\documentclass[11pt]{article}
\pdfoutput=1
\usepackage{jheppubmod}
\usepackage[punctsep]{collref}
\collectsep[]{;}
\newcommand{\bl}[1]{#1}

\def\la{\lambda}
\def\lad{\lambda^{\dagger}}
\def\lbd{\bar{\lambda}^{\dagger}}

\def\lb{\bar{\lambda}}

\def\mb{\ {\mu}}

\def\ep{\epsilon}

\def\Or[#1]{{\text{O}}\left({#1}\right)}
\def\dotl[#1,#2]{\left\langle #1,\, #2 \right\rangle}
\def\dotlb[#1,#2]{\left\langle #1,\, #2 \right\rangle}
\def\dotlm[#1,#2]{\left[ #1,\, #2 \right]}
\def\dotp[#1,#2]{(\vect{#1} \cdot\vect{#2})}
\def\aff[#1,#2]{\hat{#1}(#2)}
\def\n4sym{{\cal N}=4 SYM}
\def\>{\rangle}
\def\<{\langle}
\def\weight[#1,#2,#3]{\{(#1),#2,#3\}}
\def\ads[#1]{$\text{AdS}_{#1}$}

\newcommand{\be}{\begin{equation}}
\newcommand{\ee}{\end{equation}}
\newcommand{\bea}{\begin{eqnarray}}
\newcommand{\eea}{\end{eqnarray}}
\newcommand{\ba}{\begin{align}}
\newcommand{\ea}{\end{align}}
\newcommand{\nn}{\nonumber \\}
\newcommand{\bs}{\begin{split}}
\def\sess\end{split}
\newcommand\onlinecite{\cite}
\newcommand{\vect}[1]{{\boldsymbol{#1}}}
\newcommand{\norm}[1]{|{\boldsymbol{#1}}|}
\usepackage{hyperref}
\usepackage{setspace} 
\title{Multipoint correlators of conformal field theories:\\
implications for quantum critical transport}

\author[1]{Debanjan Chowdhury,}
\affiliation[1]{Department of Physics, Harvard University, Cambridge MA
02138, USA.}
\author[2,3,4]{Suvrat Raju,}
\affiliation[2]{International Centre for Theoretical Sciences, TIFR, IISc Campus,
  Bangalore 560012, India.}
\affiliation[3]{Harish-Chandra Research Institute, Jhunsi,
Allahabad 211019, India.}
\affiliation[4]{School of Natural Sciences, Institute for Advanced Study,
  Princeton, NJ 08540, USA.}
\author[1]{Subir Sachdev,}
\author[5,6]{Ajay Singh,}
\affiliation[5]{Department of Physics \& Astronomy and Guelph-Waterloo Physics Institute,
University of Waterloo, Waterloo, Ontario N2L 3G1, Canada.}
\affiliation[6]{Perimeter Institute for Theoretical Physics, Waterloo, Ontario N2L 2Y5, Canada.}
\author[1]{\mbox{Philipp Strack}}

\abstract{
We compute three-point correlators between the stress-energy tensor and conserved currents of conformal field
theories (CFTs) in 2+1 dimensions. We first compute the correlators in the large-flavor-number expansion of conformal gauge theories and then do the computation using holography. In the holographic approach, the correlators are computed from an effective action on 3+1 dimensional anti-de Sitter space (AdS$_4$), and depend upon the co-efficient, $\gamma$, of a four-derivative term in the action.
We find a precise match between the CFT and
the holographic results, thus fixing the values of $\gamma$. The CFTs of free fermions and bosons take the values $\gamma=1/12,-1/12$
respectively, and so saturate the bound $|\gamma| \leq 1/12$ obtained earlier from the holographic theory; the correlator of the
conserved gauge flux of 
U(1) gauge theories takes intermediate values of $\gamma$. The value of $\gamma$ also controls the frequency dependence of the
conductivity, and other properties of quantum-critical transport at non-zero temperatures. Our results for the values of $\gamma$
lead to an appealing physical interpretation of particle-like or vortex-like transport near quantum phase transitions of interest
in condensed matter physics. \bl{This paper includes appendices reviewing key features of the AdS/CFT correspondence for condensed matter
physicists.}
}

\preprint{\parbox{3cm}{ICTS/2012/07 \\ HRI/ST/1206 \\ arXiv:1210.5247}}
\begin{document}
\maketitle


\section{Introduction}

\begin{figure}[t]
\begin{center}
\includegraphics[width=6.2in]{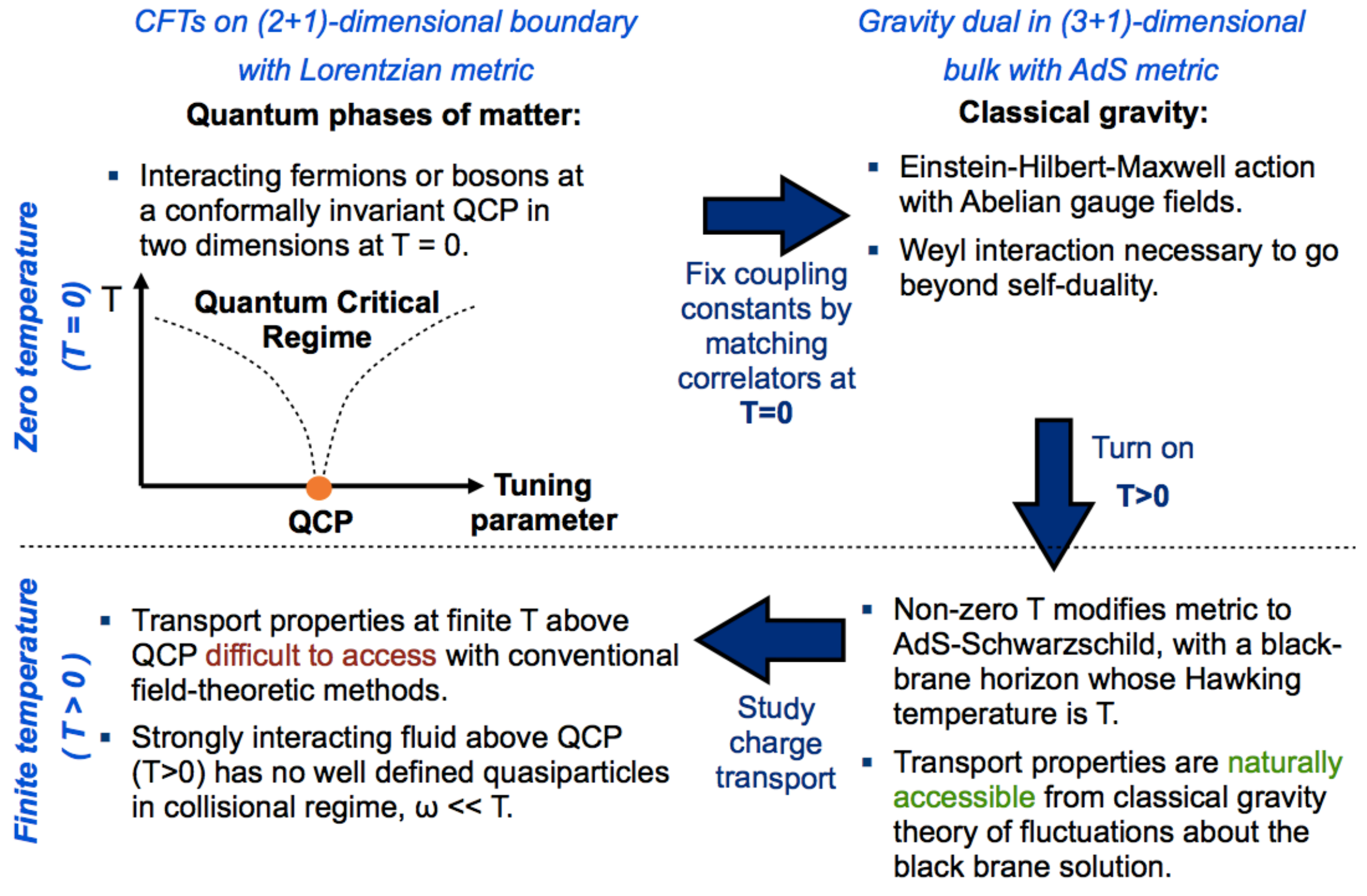}
\end{center}
\caption{Illustration of the AdS-CFT correspondence in the context of quantum critical transport at finite temperatures. The present paper is concerned with the upper blue arrow: we fix couplings by matching correlators of the CFT to those of the gravity theory. The bottom blue 
arrow is addressed in Refs.~\onlinecite{Myers:2010pk} and~\onlinecite{quasinormal}, which computed the relevant conductivities and quasi-normal modes
of the gravity dual for general values of the couplings in Eq.~(\ref{Smyers}).}
\label{fig:mapping}
\end{figure}

This paper is a contribution to the program of connecting strongly interacting condensed matter systems to 
theories based upon the methods of gauge-gravity duality \cite{Maldacena:1997re,Gubser:1998bc,Witten:1998qj}. 
Such methods offer powerful tools to describe dynamics at non-zero temperatures,
and far from equilibrium, in regimes far-removed from any quasiparticle theory. But they have been rigorously established only for strongly
interacting non-Abelian gauge theories which are very different from those relevant for condensed matter applications.
For the latter, the simplest context in which the connections may be made are
conformal field theories (CFTs) in 2+1 dimensions \cite{Herzog:2007ij}
which are dual to gravity theories on AdS$_4$. 
Myers {\em et al.} \cite{Myers:2010pk}  proposed extending the gauge-gravity methods to a wider class of CFTs by 
viewing the gravity theory as a phenomenological effective field theory \bl{on 3+1 dimensional anti-de Sitter space (AdS$_4$)}, with physical observables to be
computed in the gravity theory at {\em tree level\/}. The effective field theory
was expanded in powers of spacetime gradients, and all terms with up to 4 gradients were retained; such a field
theory was also considered earlier by Ritz and Ward \cite{Ritz:2008kh}.
In this paper, we will pin down the values some of the coupling constants in this holographic theory by a matching
procedure based upon the computation of 3-point correlators of the stress-energy tensor and the conserved currents
at zero temperature ($T$) \cite{Hofman:2008ar}. 
For the case of linear response functions of charge transport in a CFT at zero density, all needed 4 gradient couplings will be determined;
we review the arguments for this in Appendix~\ref{app:action}.
This allows us to relate CFTs of interest in condensed matter to a specific holographic action.
And it paves the way for predictions on the non-zero $T$ and non-equilibrium dynamics for condensed matter systems
from holographic methods as illustrated in Fig.~\ref{fig:mapping}.

We have written this paper for readers with a background in condensed matter theory, and a knowledge of general relativity.
Readers with no prior knowledge of gauge-gravity duality are referred
to a recent review article \cite{Sachdev:2011wg} for an overall perspective, 
and to Appendix~\ref{sec:2point} for a description of the correspondence between correlators of the CFT and
the theory on AdS$_4$.

While our results are quite general, it is useful to express them in the context of a particular CFT
which has numerous condensed matter applications
\cite{wen1993transitions,chen1993mott,rantner2001electron,sachdev1998nonzero,rantner2002spin,motrunich2004emergent,senthil2004deconfined,senthil2004quantum,hermele2004stability,hermele2005algebraic,kaul2007algebraic,kaul2008quantum,Klebanov:2011td}.
The matter sector has Dirac fermions $\psi_\alpha$, $\alpha = 1 \dots N_f$,
and complex scalars, $z_a$, $a = 1 \ldots N_s$. We will always take the large $N_f$ limit with $N_s/N_f$ fixed, and use the symbol $N_F$
to refer generically to either $N_s$ or $N_f$. These matter fields
are coupled to each other and a U(1) gauge field $a_i$ by a Lagrangian of the form
\be
\mathcal{L} = \sum_{\alpha=1}^{N_f} i \overline{\psi}_\alpha \gamma^i D_i \psi_\alpha + \sum_{a=1}^{N_s}
\bigg ( |D_i z_a|^2 + s |z_a|^2 + \frac{u}2 \left( |z_a|^2 \right)^2 \bigg ) + \ldots \,, \label{lcft}
\ee
where $D_i = \partial_i - i a_i$ is the gauge covariant derivative, the Dirac matrices obey $\mbox{Tr} ( \gamma^i \gamma^j) = 2 \eta^{ij}$ where $\eta^{ij}$ is the 
Minkowski metric, and the ellipses represent additional possible contact-couplings
between the fermions and bosons. The scalar ``mass'' term, $s$, has to be tuned to reach
the quantum critical point, which is described by a CFT at the renormalization group (RG) fixed point; fermion mass terms can be removed
by imposing discrete symmetries. So the scalar mass is the only relevant perturbation at the CFT fixed point 
and only a single parameter has to be tuned to access the fixed point. All other couplings, such as $u$ and the Yukawa coupling, reach values associated with the RG fixed point, and so their values are immaterial for the universal properties of interest in the present paper.

This CFT has three globally conserved currents. There is the SU($N_s$) scalar flavor current
\be
J_{s, i}^\ell = - i z_a^\ast \, T^\ell_{ab} \, \left(D_i z_b \right) + i \left(D_i z_a\right)^\ast \, T^{\ell}_{ab} \, z_b \label{Js} , 
\ee
where $T^\ell$ are the generators of SU($N_s$) normalized by $\mbox{Tr} ( T^\ell T^m) = \delta^{\ell m}$.
Similarly there is 
the fermion SU($N_f$) flavor current
\be
J_{f, i}^\ell =  \overline{\psi}_{\alpha} \, T^\ell_{\alpha\beta} \gamma_i \, \psi_\beta. \label{Jf}
\ee
Finally, there is the topological U(1) current
\be
J_{t, i} = \frac{1}{2 \pi} \epsilon_{ijk} \partial^j a^k. \label{Jt}
\ee
We will use the symbol $J_i$ to generically refer to any one of these three currents. A basic property of the CFT \cite{Osborn:1993cr}
is that the two-point
correlator of a conserved current obeys
\be
\left\langle J_{i} (\vect{k})  J_{j} (- \vect{k}) \right\rangle = - C_J \, |\vect{k}| \left( \eta_{ij} - \frac{k_i k_j}{|\vect{k}|^2} \right) , \label{ej}
\ee
where $\vect{k}$ is a spacetime momentum, $\eta_{ij} = {\rm diag} (-1,1,1)$ is the Minkowski metric, and $C_J$ is a dimensionless universal constant associated with the CFT and the current. Similarly, the stress-energy tensor, $T_{ij}$, of the CFT has the two-point correlator \cite{Cardy1987355}
\bea
\left\langle T_{ij} (\vect{k} ) T_{u v} (-\vect{k} ) \right\rangle 
&=& C_T |\vect{k}|^3 \left( \eta_{iu} \eta_{j v} + \eta_{ju} \eta_{i v} - \eta_{ij} \eta_{uv} + \eta_{ij} \frac{k_u k_v}{|\vect{k}|^2} + \eta_{u v} \frac{k_i k_j}{|\vect{k}|^2} \right. \nn
&~& \left. - \eta_{iu} \frac{k_j k_v}{|\vect{k}|^2} - \eta_{ju} \frac{k_i k_v}{|\vect{k}|^2}- \eta_{iv} \frac{k_j k_u}{|\vect{k}|^2}- \eta_{jv} \frac{k_i k_u}{|\vect{k}|^2}
+ \frac{k_i k_j k_u k_v}{|\vect{k}|^4} \right) , \label{et}
\eea
where $C_T$ is another universal constant characterizing the CFT.

The primary focus of the present paper will be on the structure of the 3-point correlator 
$\langle T_{i_1 j_1}(\vect{k_1}) J_{i_2}(\vect{k_2}) J_{i_3}(\vect{k_3}) \rangle$. The general form of this correlator
for a CFT was specified by Osborn and Petkou \cite{Osborn:1993cr} in position space: they showed that it was fully determined by the values of 
$C_J$, $C_T$, and a single additional constant. Such a position space correlator was matched to holographic results by 
Hofman and Maldacena \cite{Hofman:2008ar}, and we will follow their methods in Section~\ref{sec:ajay}.
However, we will first perform this computation in momentum space.
It is not a simple matter to take the Fourier transform of the earlier position space result \cite{Osborn:1993cr},
and we will therefore compute this correlator
directly from the CFT, and from its holographic partner. 

\begin{figure}[t]
\begin{center}
\includegraphics[width=50mm]{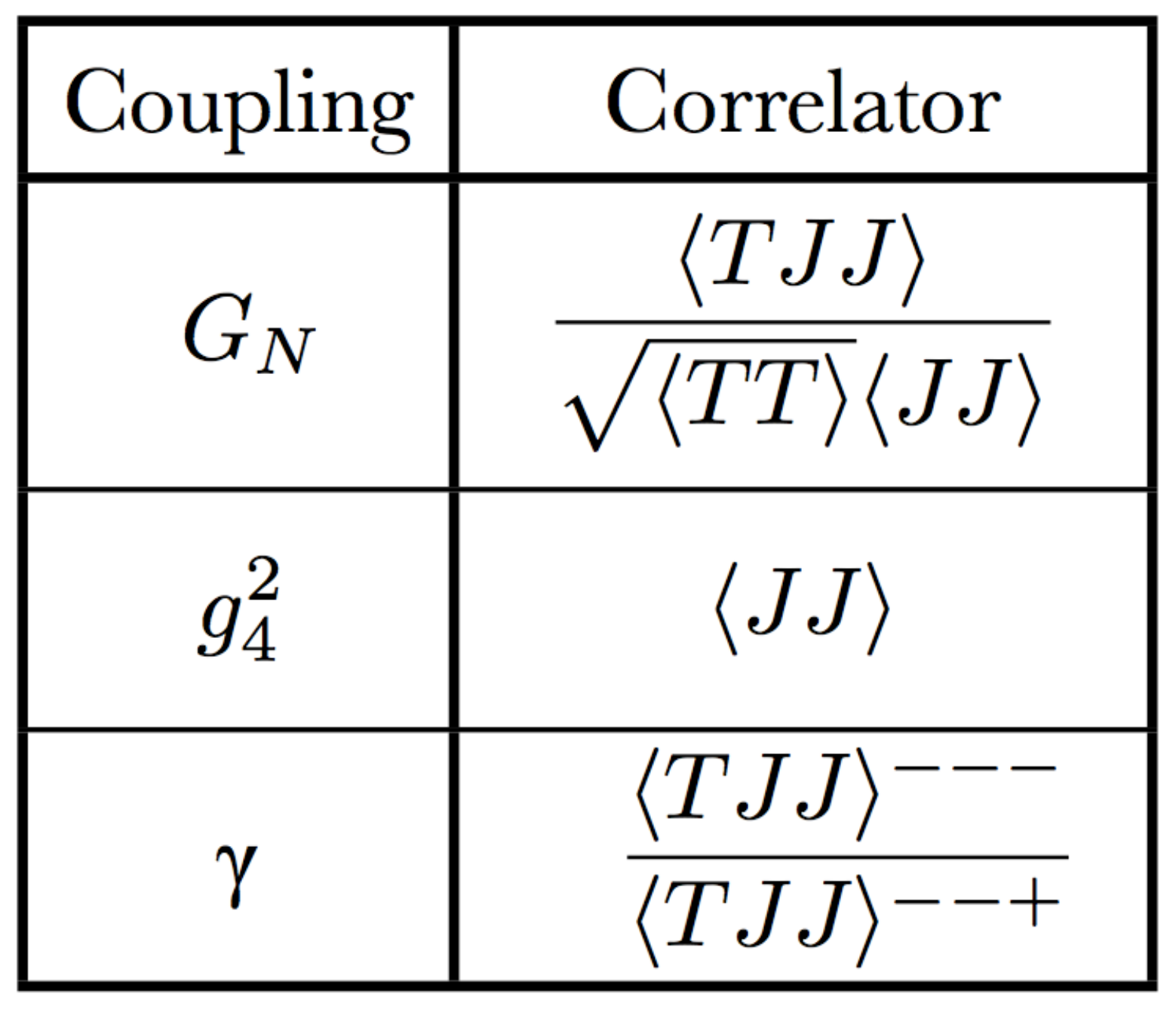}
\end{center}
\caption{Correlators (with helicity projections) 
that fix the numerical values of the couplings in the holographic action specified by 
Eqs.~(\ref{Smyers}) and (\ref{heaction}). These correlators are evaluated in the present paper in the boundary
conformal field theory.}
\label{fig:couplings}
\end{figure}

Our purpose is to relate the conserved current correlators of the CFT (\ref{lcft}) to the effective holographic theory
of Refs.~\cite{Myers:2010pk,Ritz:2008kh}. The theory is defined on AdS$_4$, and has a (non-Abelian or Abelian) 
gauge field $A_\mu$, and corresponding gauge flux $F_{\mu\nu}$,
associated with each of the conserved currents $J_i$. (Our convention is that Greek indices run over all directions in the bulk, while Latin indices are used to denote boundary directions.)
We note that there is no direct relationship between the bulk gauge field $A_\mu$ and the boundary gauge field $a_i$.
As we review in Appendix~\ref{app:action}, the most general 4-derivative action for linear transport of the CFT in each bulk gauge field is
\be
S = 
{1 \over g_4^2} 
\int d^4 x \sqrt{-g}\, \mbox{Tr} \, \left[-{1 \over 4} F_{\mu \nu} F^{\mu\nu}  + \gamma L^2 \, C_{\mu \nu \rho \sigma} 
F^{\mu\nu} F^{\rho\sigma} \right], \label{Smyers}
\ee
where $\mbox{Tr}$ is the trace over $SU(N_s)$ or $SU(N_f)$ indices (if present), 
$g_{\mu\nu}$ is the metric of AdS$_4$ with radius $L$, 
and $C_{\mu\nu\rho\sigma}$ is the Weyl curvature tensor; we will usually set $L=1$, although it will be reinstated in 
some final results. We reiterate the conditions under which \eqref{Smyers} constitutes the most general tree-level effective holographic theory: for linear charge transport in CFTs in the absence of a chemical potential.
As we will review below, matching the 
two-point correlator of the current between (\ref{lcft}) and (\ref{Smyers}) fixes the value of the coupling $g_4$. The coupling crucial for our
purposes is $\gamma$; it was shown that the stability of the theory $S$ requires $|\gamma | \leq 1/12$.
The structure of the 3-point correlator 
$\langle T_{i_1 j_1}(\vect{k_1}) J_{i_2}(\vect{k_2}) J_{i_3}(\vect{k_3}) \rangle$ is determined by $\gamma$, and so $\gamma$
will play the role of the additional constant noted by Osborn and
Petkou \cite{Osborn:1993cr} (the explicit relation to their constants is specified in Section~\ref{sec:ajay}).
Comparison with the CFT computation
yields the value of $\gamma$. An overview of the correlation functions needed to fix the values of 
the coupling constants in Eq.~(\ref{Smyers}) is given in Fig.~\ref{fig:couplings}.

Our results for the values of $\gamma$ for the currents in (\ref{Js}), (\ref{Jf}), and (\ref{Jt}) are
\bea
\gamma_f &=& \frac{1}{12} + \mathcal{O}(1/N_F), \nn 
\gamma_s &=& - \frac{1}{12} + \mathcal{O}(1/N_F), \nn  
\gamma_t &=& \frac{N_s - N_f}{12 (N_s + N_f)} + \mathcal{O}(1/N_F). \label{gammares}
\eea
It is interesting that the free CFT results ($\gamma_f$ and $\gamma_s$ at $N_F = \infty$) 
saturate the bound on $\gamma$ in the large $N_F$ limit. 
We recall that a similar feature was observed in earlier computations of three-point correlators of the stress energy tensor,
where the free field results also saturate the bounds obtained from the holographic higher derivative theory \cite{Buchel:2009sk,Myers:2010jv}.

For $N_f=0$ we have $\gamma_s = - \gamma_t$. This change in sign of $\gamma$ is consistent with
the expectations \cite{Myers:2010pk} of its transformation under particle-vortex duality, and the interpretation of $J_{ti}$ as the matter current in the dual theory.
Further discussion on the physical consequences of these values of $\gamma$ appear in 
Section~\ref{sec:conc}.

We note that 3-point correlators of CFTs have also played an important role in recent investigations
of theories with higher-spin conserved currents
\cite{Giombi:2012he,
  Giombi:2011kc,Maldacena:2011jn,Aharony:2012nh}. 
Our 3-point correlator is similar, but our holographic considerations
follow a different route.

The outline of the rest of the paper is as follows. In
Section~\ref{sec:setting} we describe the setting, in which we will
perform our correlation function calculations. Section~\ref{sec:CFT3} will present the computation of the 3-point correlator in the large $N_F$ limit of the CFT.
In Section~\ref{sec:holo3} we will present the holographic computation of the 3-point correlator implied by  the
AdS$_4$ action of Myers {\em et al.} at tree level.  The two sets of results will be matched
in Section~\ref{sec:matching}. 
Section~\ref{sec:ajay} will present another derivation of our values of $\gamma$ for the free field theories,
using the methods of Ref.~\onlinecite{Hofman:2008ar}.
In Section~\ref{sec:conc}, we
explore some of the consequences of these results.

\section{Setting \label{sec:setting}}
In this section, we introduce our momentum space notation for the 3-point correlators and briefly recapitulate the 
spinor-helicity projections that we perform in the CFT as well as in the holography computation. 
The momentum space expressions of the 3-point correlator are obtained by Fourier transforming along the boundary
directions:
\be
\label{undottedcorr}
\begin{split}
\mathcal{K}^{i_1 j_1 i_2 i_3} (\vect{k}_1, \vect{k}_2 , \vect{k}_3 )
&=  \left\langle T^{i_1 j_1}(\vect{k_1}) J^{i_2}(\vect{k_2})
J^{i_3}(\vect{k_3}) \right\rangle \\ &\equiv  \int \left\langle {\cal
  T}\Big\{T^{i_1 j_1}(\vect{x_1}) J^{i_2}(\vect{x_2})
J^{i_3}(\vect{x_3}) \Big\} \right\rangle e^{i \sum_{m=1}^3\vect{k_m} \cdot
  \vect{x_m}} d^3 x_m, 
\end{split}
\ee
where ${\cal T}$ is the time-ordering symbol and the integral runs
over the three flat directions on the boundary. (The time-ordered
correlator is what we would get by computing the Euclidean
correlation function and then Wick-rotating to Lorentzian space.)

There are several advantages of working in momentum space.  Some of
these will become apparent below, but let us comment on one immediate
benefit. In \ref{undottedcorr}, we have many free indices. In
particular, for the stress-tensor, the Ward identities tell us that if
we contract $i_1$ and $j_1$ in \eqref{undottedcorr}, this will yield
a known answer in terms of lower point correlators. Similarly
\eqref{undottedcorr} is symmetric in $i_1,j_1$ up to contact terms
that again involve two-point functions. However, this still naively
leaves us with $5$ degrees of freedom in the stress-tensor and $3$ in
each of the currents. 

However, both the stress-tensor and the currents are conserved. In
position space, this leads to differential Ward identities. In
momentum space, these identities become algebraic: they translate to
the simple statement that, for $m=1,2,3$,  the contraction $k_{m, i_m} {\cal
  K}^{i_1j_1 i_2 i_3}$ is determined in terms of lower-point
correlators.  

This means that we can extract all the physical information in
\eqref{undottedcorr} by contracting the stress-tensor with any two
symmetric and traceless polarization tensors that are transverse to the
momentum $\vect{k_1}$, and the two currents with polarization vectors that
are transverse to $\vect{k_2}$ and $\vect{k_3}$ respectively. So, we
can instead consider
\be
\label{dottedcorr}
\mathcal{K} (\vect{e_1}, \vect{k_1},\vect{\ep_2}, \vect{k_2},\vect{\ep_3}, \vect{k_3}) = e_{1, i_1 j_1} \ep_{2, i_2} \ep_{3, i_3} \langle T^{i_1 j_1}(\vect{k_1}) J^{i_2}(\vect{k_2}) J^{i_3}(\vect{k_3}) \rangle.
\ee
Here $\vect{e_1}$ is a polarization-tensor for the stress-tensor, and $\vect{\ep_2}$ and $\vect{\ep_3}$ are polarization vectors for the currents. We choose these to
be transverse to the momentum carried by the
corresponding operator, and it will also be convenient for us to
choose them to be null:
\be
\vect{\ep_m} \cdot \vect{k_m} = \vect{\ep_m} \cdot \vect{\ep_m} = 0.
\ee
We can choose the
polarization tensor $\vect{e_1}$ to be just an outer product of two polarization
vectors for $\vect{k_1}$:
\be
e_{1, i j} = \ep_{1, i} \ep_{1, j}.
\ee

So, the use of momentum space drastically cuts down the number of
independent indices that we need to deal with and allows us to
directly engage with the physical quantities in \eqref{dottedcorr}. 

To simplify the algebra even further, we will use the {\it spinor-helicity}
formalism to write down explicit expressions for the polarization
tensors and, later, to simplify the correlators.   The spinor-helicity
formalism was initially introduced to study four dimensional
scattering amplitudes, as a means of efficiently encoding the
kinematics of the external particles. (See Ref.~\onlinecite{Dixon:1996wi} and
references there.) It was adapted to the study of correlators in
three-dimensional conformal field theories by Maldacena and Pimentel
\cite{Maldacena:2011nz}. 

Our conventions are different from those of Ref.~\onlinecite{Maldacena:2011nz},
and we provide a detailed introduction to this formalism in Appendix
\ref{secspinoreview}. Here, we excerpt a few of the essential
details to help the reader parse the formulas in this paper.

Given a three vector $\vect{k} = (k_0, k_1, k_2)$, we consider the $2
\times 2$ matrix
\begin{equation}
k_{\alpha \dot{\alpha}} = k_0 \sigma^0_{\alpha \dot{\alpha}} +  k_1
\sigma^1_{\alpha \dot{\alpha}} +  k_2 \sigma^2_{\alpha \dot{\alpha}} +
i |\vect{k}| \sigma^3_{\alpha \dot{\alpha}},
\end{equation}
where $|\vect{k}| \equiv \sqrt{\vect{k} \cdot \vect{k}} = \sqrt{k_1^2
  + k_2^2 - k_0^2}$. By construction, this $2 \times 2$ matrix has
rank $1$ and so it can be decomposed into the outer product of a $2
\times 1$ and a $1 \times 2$ vector:
\be
k_{\alpha \dot{\alpha}} = \la_{\alpha} \lb_{\dot{\alpha}}.
\ee
The $\la$ and $\lb$ are called spinors, and instead of giving the
momentum 3-vectors for each operator insertion, we can instead give
these spinors. 

We can define dot products of these spinors via: 
\be
\label{spinordot}
\dotl[\la_1, \la_2] = \epsilon^{\alpha \beta} \la_{1 \alpha} \la_{2 \beta} = \la_{1 \alpha} \la_{2}^{\alpha}, \quad \dotl[\lb_1, \lb_2] = \epsilon^{\dot{\alpha} \dot{\beta}} \lb_{1 \dot{\alpha}} \lb_{2 \dot{\beta}} = \lb_{1 \dot{\alpha}} \lb_2^{\dot{\alpha}}. 
\ee
Finally, one other advantage of this formalism is that the
polarization vectors we referred to above can be written quite
easily in terms of these spinors:
\begin{equation}
\label{polarizationvects}
\epsilon^+_{\alpha \dot{\alpha}} = 2 {\lbd_{\alpha} \lb_{\dot{\alpha}}
  \over \dotlm[ \la, \lb]} =  {\lbd_{\alpha} \lb_{\dot{\alpha}} \over
  i \norm{k}}, \quad \epsilon^-_{\alpha \dot{\alpha}} =  2 {\la_{\alpha} \lad_{\dot{\alpha}} \over \dotlm[\la, \lb]} = {\la_{\alpha} \lad_{\dot{\alpha}} \over i \norm{k}},
\end{equation}
where we have labeled the polarization vectors by a ``helicity'' that
can either be positive or negative.  We refer the interested reader
to Appendix \ref{secspinoreview} for further details.

\section{CFT computation of 3-point correlators}
\label{sec:CFT3}

In this section, we compute the three-point correlators of each of the conserved currents (\ref{Js}), (\ref{Jf}), and (\ref{Jt}) 
for the Lagrangian (\ref{lcft}) with its couplings at the CFT fixed point.
The stress-energy tensor is
\be
T_{ij} = T_{s,ij} + T_{f,ij},
\ee
which consists of a scalar, bosonic contribution
\be
T_{s,ij} = \sum_{a=1}^{N_s} \bigg ( \left(D_i z_a \right)^\ast\left( D_j z_a \right) 
+ \left(D_j z_a \right)^\ast \left( D_i z_a \right) - \frac{1}{4} \left( \partial_i \partial_j + \eta_{ij} \partial^2 \right) |z_a|^2 \bigg )  ,
\ee
and the fermionic contribution
\be
T_{f,ij} = \frac{i}{4} \sum_{\alpha=1}^{N_f} \bigg( \overline{\psi}_{\alpha} \gamma_i \left( D_j \psi_\alpha \right) +  \overline{\psi}_{\alpha} \gamma_j \left( D_i \psi_\alpha \right) -  \left( D_i^\ast \overline{\psi}_{\alpha} \right) \gamma_j  \psi_\alpha -  \left( D_j^\ast \overline{\psi}_{\alpha} \right) \gamma_i  \psi_\alpha 
\bigg).
\ee
We evaluate the correlators by summing over all possible Wick contractions of the constituent operators of 
$\langle T J J \rangle$ defined in (\ref{undottedcorr}) in the limit of large flavor number $N_F$. As expected, we will see that the leading contractions with the flavor currents are those of the free CFT. For the topological currents the first non-vanishing contractions appear at $\mathcal{O}(1/N_F)$. 
All contractions involve tensor-valued one-loop integrations in momentum space which we evaluate 
using Davydychev recursion relations \cite{Davydychev:1991va,Davydychev:1992xr,Bzowski:2011ab}. Finally, the full 
tensor-valued expressions are contracted with the polarization or helicity operators defined in Sec.~\ref{sec:setting} 
to bring them to a form that facilitates comparison with the corresponding helicity projections from the holographic calculation 
(performed in Sec.~\ref{sec:holo3}).

We refer the readers to Appendix~\ref{app:two_point} for a review of the computations of the two-point functions, $\langle JJ \rangle$ and $\langle T T\rangle$,  leading to (\ref{ej}, \ref{et}) and the final results after contracting with the corresponding polarization tensors. 

\subsection{$\langle T J J \rangle$ for SU$(N_s)$ scalar flavor current}
\label{subsec:sca}

\begin{figure}[h]
\begin{center}
\includegraphics[width=5in]{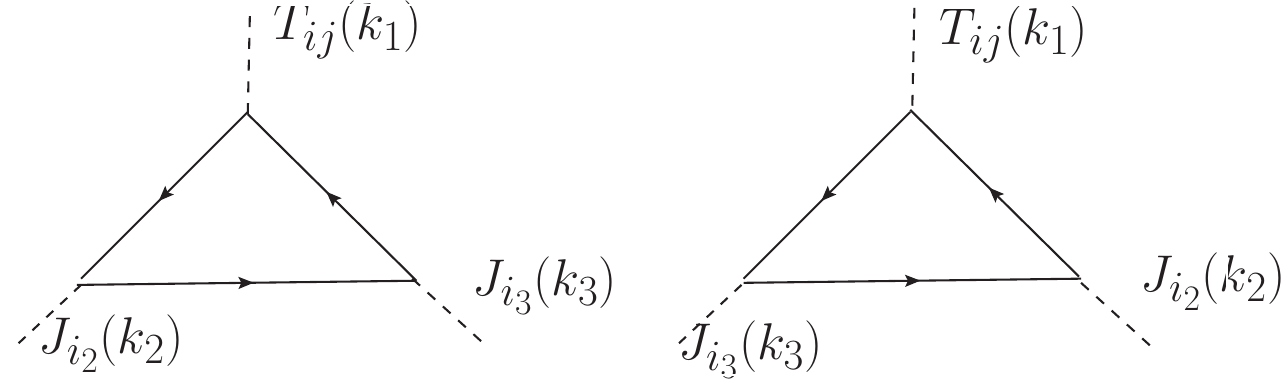}
\end{center}
\caption{One-loop triangle diagrams for the scalar contribution to $\langle T J J \rangle$. The top corner of the respective triangles are (momentum-dependent) stress-tensor vertices while the bottom two corners represent current vertices.}
\label{fig:diagrams}
\end{figure}

Evaluating Wick's theorem for the scalar correlator yields two non-vanishing contractions depicted diagrammatically 
in Fig.~\ref{fig:diagrams}.
The full expression for the two diagrams is:
\be
\label{eq:tjj_scalar}
\begin{split}
\mathcal{K}_s^{i_1j_1i_2i_3}(\vect{ k_1},\vect{ k_2},\vect{ k_3})=\int &\frac{d^3 P}{8 \pi^3} \frac{4}{P^2 (\vect{ P} + \vect{ k_1})^2 (\vect{ P} - \vect{ k_2})^2} ( 2 \vect{ P} - \vect{ k_{2}})^{i_2} ( 2 \vect{ P} + \vect{ k_{1}})^{i_3}
 \\
&\times \left[ \frac{1}{2} (\vect{ P} - \vect{ k_{2}})^{i_1}(\vect{ P} + \vect{ k_{1}})^{j_1} + \frac{1}{2} (\vect{ P} - \vect{k_{2}})^{j_1}(\vect{ P} + \vect{ k_{1}})^{i_1} \right. \\ &\left.+ \frac{1}{8} ( \norm{k_3}^2 \eta^{i_1j_1} + (\vect{ k_{1}} + \vect{ k_{2}})^{i_1}(\vect{ k_{1}} + \vect{k_{2}})^{j_1}) \right],
\end{split}
\ee
with ${\bf k_1}+{\bf k_2}+{\bf k_3}=0$. The momentum dependence in the numerator of (\ref{eq:tjj_scalar}) comes from derivative operators of the fields at each vertex. We are only interested in certain polarization projections of this expression and we now explain how this simplifies the momentum structure considerably.

\begin{figure}
\begin{center}
 \includegraphics[width=3in]{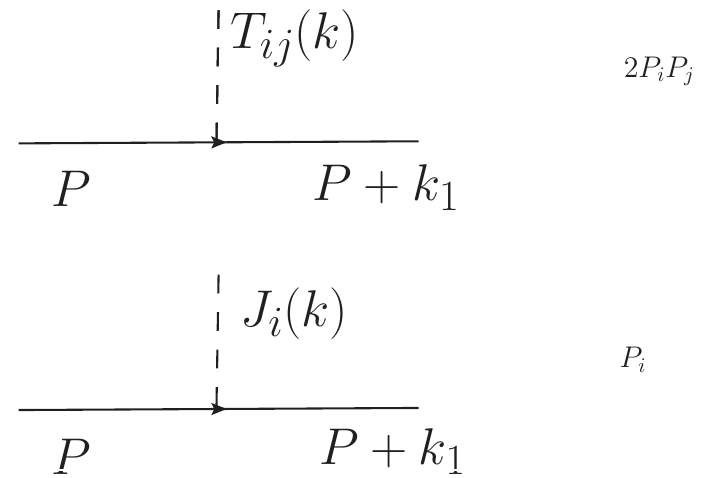}
 \caption{Momentum structure of the stress tensor (top) and current vertex (bottom) after contracting with transverse and traceless polarization tensors.}
\label{fig:vertices}
\end{center}
\end{figure}

Quite generally, a current insertion with momentum $\vect{k}$ at a vertex where one line brings in $\vect{P}$ (Fig. \ref{fig:vertices}) and the other line carries away $\vect{P} + \vect{k}$ leads to an effective vertex:
$\left(2 P_i + k_i \right)$.
However, since this correlator will finally be dotted with a transverse polarization vector, one can drop the $k_i$ term on the right hand side 
in the computations below.
Also, here and below we have dropped the SU($N_s$) generator $T^\ell$ because it only yields factors of unity after tracing over SU($N_s$) indices.
Similarly, a stress-tensor insertion carrying momentum $\vect{k}$ at a 
vertex where one line brings in the loop-momentum $\vect{P}$ (Fig. \ref{fig:vertices}) and the other line carries away $\vect{P}+\vect{k}$ results 
in a vertex that we are finally going to contract with a polarization tensor that is {\em transverse and traceless}. Since this tensor will satisfy $e^{i j} \eta_{i j} = 0 = e^{i j} k_i$, we can drop the terms proportional to $\eta_{i j}$ and also the terms proportional to $k_i$ and $k_j$ above. Using this logic, the expressions for the effective stress tensor and current vertex, respectively, become quite simple (see Fig.~\ref{fig:vertices}) and from Eq.~(\ref{eq:tjj_scalar}) we only need to consider
\be
8 N_s e_{i_1 j_1} \ep_{i_2} \ep_{i_3} \int {d^3 P \over 8 \pi^3} \left[{P_{i_1} P_{j_1}  (\vect{P}+\vect{k_1})_{i_2}  P_{i_3} \over P^2 (\vect{P}+ \vect{k_1})^2 (\vect{P} - \vect{k_3})^2} + {P_{i_1} P_{j_1} (\vect{P}+\vect{k_1})_{i_3} P_{i_2} \over P^2 ( \vect{P} + \vect{k_1})^2 (\vect{P} - \vect{k_2})^2} \right].
\ee
These integrals can be done by automating the Davydychev recursion relations \cite{Davydychev:1991va,Davydychev:1992xr,Bzowski:2011ab}. The resulting expressions are quite lengthy, as shown in the attached Mathematica file \cite{simplifyscalars}. However, after we dot this answer with polarization tensors and rewrite it using the spinor helicity formalism, our final answers are quite simple. The interested reader should again consult the Mathematica file for details. We find the 
following results for $N_s$ complex scalars, which we will later compare to the results obtained from holography:
\be
\label{freepmm}
\begin{split}
{1 \over N_s} \mathcal{K}^{+--}_{s} &=  {-\dotl[\la_3,\la_2]^4 \over 32  \dotl[\la_2,\la_1]^2  \dotl[\la_3,\la_1]^2\norm{k_{1}}^2 \norm{k_{2}} \norm{k_{3}}}
\left[\left(\norm{k_{1}}^2-(\norm{k_{2}}-\norm{k_{3}})^2\right)^2 (\norm{k_{2}}+\norm{k_{3}})\right].
\end{split}
\ee

Contracting the stress tensor with a negative helicity polarization
tensor and both the currents with negative helicity polarization
vectors leads to:
\be
\label{freemmm}
{1 \over N_s} \mathcal{K}^{---}_{s} = \frac{\dotl[\la_2, \la_1]^2 \dotl[\la_3, \la_1]^2}{32 \norm{k_{1}}^2}{ \left(\frac{8
   \norm{k_{1}}^3}{(\norm{k_{1}}+\norm{k_{2}}+\norm{k_{3}})^4}-\frac{1}{\norm{k_{2}}}-\frac{1}{\norm{k_{3}}}\right)}.
\ee

Contracting with a negative helicity for the stress tensor and one of the currents, and a
positive helicity for the second current, we find:
\be
\label{mmpfree}
\begin{split}
{1 \over N_s} \mathcal{K}^{--+}_{s} &= \frac{\dotl[\la_2, \la_1]^4 (-\norm{k_{1}}+\norm{k_{2}}+\norm{k_{3}})^2  
} {32
   \dotl[\la_3, \la_2]^2  \norm{k_{1}}^2 \norm{k_{2}} \norm{k_{3}} (\norm{k_{1}}+\norm{k_{2}}+\norm{k_{3}})^2} 
\\ &\times \left[(\norm{k_{2}}+\norm{k_{3}}) \left(\norm{k_{1}}^2+\norm{k_{2}}^2+\norm{k_{3}}^2\right) + 2
   \left(\norm{k_{2}}^2+\norm{k_{3}}^2\right) \norm{k_{1}} \right].
\end{split}
\ee

It is worthwhile to point out that all the answers above have the correct Lorentz transformation properties on the boundary and have the correct dimension. They are also symmetric in particles $2$ and $3$ when those particles have the same helicity.
\subsection{$\langle T JJ \rangle$ for SU$(N_f)$ fermion flavor current}
\label{subsec:fer}
Now, we turn to the computation of the three-point correlator $\mathcal{K}_f$ for the fermion current $J_f$. The non-vanishing 
contractions from Wick's theorem are again given by Fig.~\ref{fig:diagrams} with fermion loop propagators and the current and stress tensor 
vertices carrying additional Dirac matrix structure instead of derivative operators, as was the case for scalars. The full expression for $\mathcal{K}_f$ is given by,
\bea
\mathcal{K}_f^{i_1j_1i_2i_3}({\bf k_1},{\bf k_2},{\bf k_3})=&-&\frac{1}{4}\bigg[\Upsilon^{i_1u_3 i_2 v_2 i_3u_2}\eta^{j_1 v_3}+i_1\leftrightarrow j_1\bigg] \nonumber\\
&&\int\frac{d^3 P}{8\pi^3}\frac{(2{\bf P}+{\bf k_{1}} - {\bf k_{2}})_{v_3} ({\bf P}-{\bf k_{2}})_{u_3} P_{v_2} ({\bf P}+{\bf k_{1}})_{u_2}}{P^2 ({\bf P} + {\bf k_1})^2 ({\bf P} - {\bf k_2})^2},\nonumber\\
\eea
with a trace over six Dirac matrices given by
\bea
\Upsilon^{i_1u_3i_2v_2i_3u_2}= 2 \, \textnormal{Tr}[\gamma^{i_1}\gamma^{u_3}\gamma^{i_2}\gamma^{v_2}\gamma^{i_3}\gamma^{u_2}].
\eea
Again the momentum integral can be done using the  Davydychev recursion relations  and the trace over Dirac matrices can be carried out using standard identities of the Clifford algebra. After contracting with polarization vectors --- the reader should consult the attached Mathematica file \cite{simplifyfermions} for details --- and simplifying further, we get:

\begin{align}
\label{freefermpmm}
{1 \over N_f} \mathcal{K}^{+--}_{f} &=  {-\dotl[\la_3,\la_2]^4 \over 64  \dotl[\la_2,\la_1]^2  \dotl[\la_3,\la_1]^2\norm{k_{1}}^2 \norm{k_{2}} \norm{k_{3}}}
\left[\left(\norm{k_{1}}^2-(\norm{k_{2}}-\norm{k_{3}})^2\right)^2
  (\norm{k_{2}}+\norm{k_{3}})\right], \\
\label{freefermmmm}
{1 \over N_f} \mathcal{K}^{---}_{f} &= \frac{\dotl[\la_2, \la_1]^2 \dotl[\la_3, \la_1]^2}{64 \norm{k_{1}}^2}{ \left(\frac{-16
   \norm{k_{1}}^3}{(\norm{k_{1}}+\norm{k_{2}}+\norm{k_{3}})^4}-\frac{1}{\norm{k_{2}}}-\frac{1}{\norm{k_{3}}}\right)}, \\
\label{mmpfreeferm}
{1 \over N_f} \mathcal{K}^{--+}_{f} &= \frac{\dotl[\la_2, \la_1]^4  (-\norm{k_{1}}+\norm{k_{2}}+\norm{k_{3}})^2}{64
    \dotl[\la_3, \la_2]^2  \norm{k_{1}}^2 \norm{k_{2}} \norm{k_{3}}
    (\norm{k_{1}}+\norm{k_{2}}+\norm{k_{3}})^2} \\ \nonumber
&\times  \left[(\norm{k_{2}}+\norm{k_{3}})
   \norm{k_{1}}^2+ (\norm{k_{2}}-\norm{k_{3}})^2 \left(2 \norm{k_{1}}+ \norm{k_{2}}+\norm{k_{3}}\right) \right]\;,
\end{align}
where we used the same conventions for the helicity superscripts as in the scalar case (\ref{freepmm}).

\subsection{$\langle T JJ \rangle$ for U$(1)$ topological current}
\label{sec:topo}

The contractions involving two topological currents (\ref{Jt}) necessarily involve two gauge field insertions and the leading 
diagrams of the $1/N_F$ expansion are shown in Fig.~\ref{fig:topo}. Although there is no bare dynamics in the gauge sector of Eq.~(\ref{lcft}), the gauge field picks an order $1/N_F$ dynamical renormalization 
from fluctuations of the scalars and fermions \cite{kaul2008quantum}, and takes the well known ``overdamped'' form:
\bea
D_{u_2v_2}({\bf q})=\langle a_{u_2} a_{v_2}\rangle=\frac{16}{(N_s+N_f)}\frac{1}{|{\bf q}|}\bigg(\eta_{u_2v_2}-\zeta\frac{q_{u_2}q_{v_2}}{{\bf q}^2} \bigg),
\eea  
where $\zeta$ is a gauge-fixing parameter that should not appear in the expression for any physical observable. With this gauge propagator, 
the diagrams in Fig.~\ref{fig:topo} evaluate to the expressions:
\begin{figure}
\begin{center}
 \includegraphics[width=3.5in]{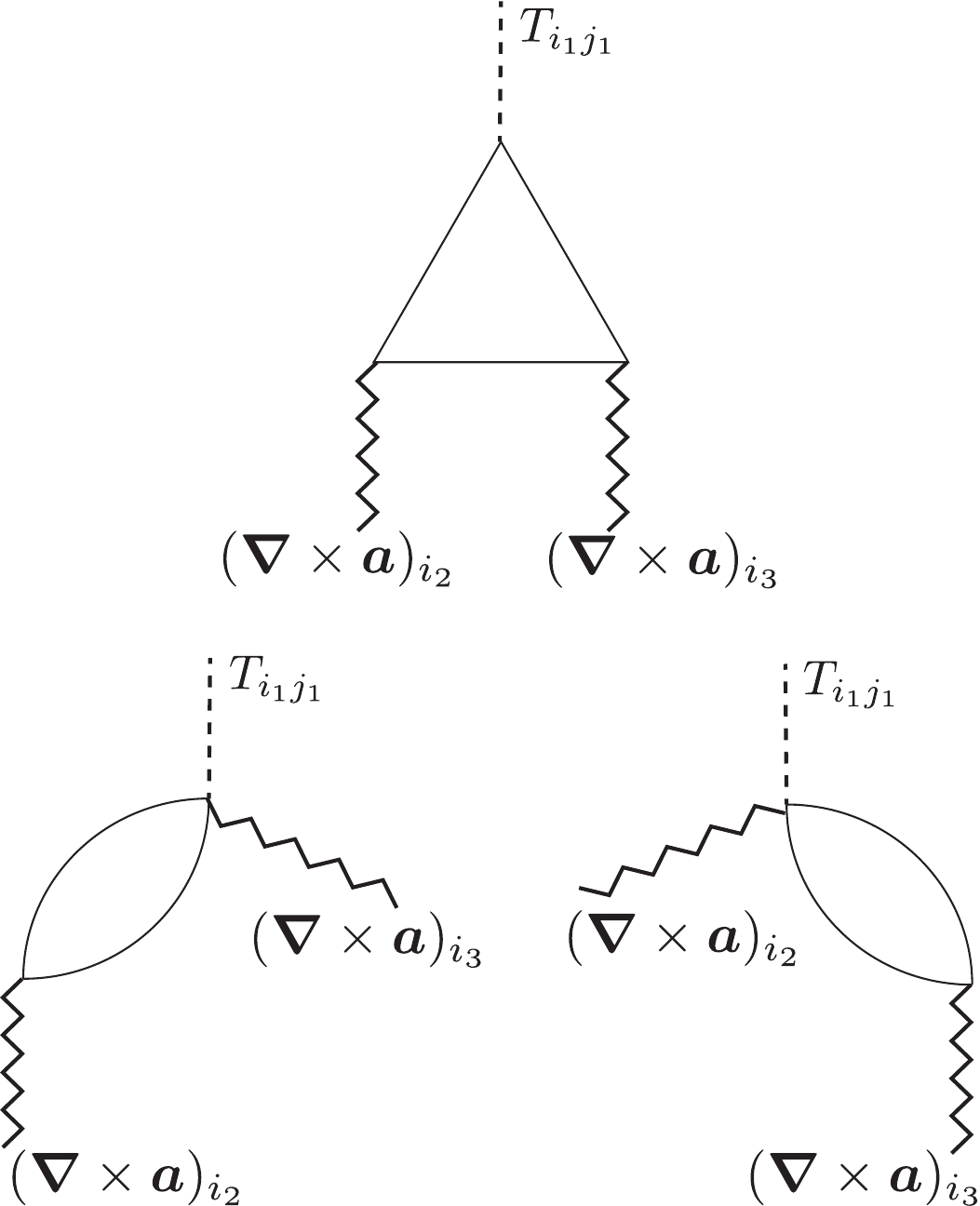}
 \caption{Feynman diagrams contributing to the 3-point correlator of $J_t$. The full lines are the bosonic or fermionic matter
 fields, and the zigzag line is the $a_i$ propagator.}
\label{fig:topo}
\end{center}
\end{figure}
\bea 
\label{ktans}
&& \mathcal{K}^{i_1 j_1 i_2 i_3}_t (\vect{k}_1, \vect{k}_2 , \vect{k}_3 ) =  \left( \frac{8}{\pi(N_f+N_s)} \right)^2 \epsilon^{i_2}_{u_2 v_2} \epsilon^{i_3}_{u_3 v_3} \, \frac{ k_{2}^{u_2} k_{3}^{u_3} }
{|\vect{k_2}| |\vect{k_3}|}   \nn
&~& \quad \times \Bigg[ N_f \Bigg \{
\mathcal{K}^{i_1 j_1 v_2 v_3}_f (\vect{k}_1, \vect{k}_2 , \vect{k}_3 ) \nn
&~& \quad \quad \quad \quad  + \frac{|\vect{k_3}|}{32} \left[ \eta^{v_2 j_1} \eta^{i_1 v_3} + \eta^{v_2 i_1} \eta^{j_1 v_3} \right]  
+ \frac{|\vect{k_2}|}{32} \left[ \eta^{v_3 j_1} \eta^{i_1 v_2}  + \eta^{v_3 i_1}  \eta^{j_1 v_2} \right] \Bigg \} \nn
&~& \quad \quad +  N_s \Bigg \{ \mathcal{K}^{i_1 j_1 v_2 v_3}_s (\vect{k}_1, \vect{k}_2 , \vect{k}_3 ) \nn
&~& \quad \quad \quad \quad  + \frac{|\vect{k_2}|}{16} \left[ \eta^{v_2 j_1} \eta^{i_1 v_3} + \eta^{v_2 i_1} \eta^{j_1 v_3} \right]  
+ \frac{|\vect{k_3}|}{16} \left[ \eta^{v_3 j_1} \eta^{i_1 v_2}  + \eta^{v_3 i_1}  \eta^{j_1 v_2} \right] \Bigg\} \Bigg]\;,
\eea
where the terms proportional to $\mathcal{K}_s$ and $\mathcal{K}_f$, respectively, originate from the top diagram in Fig.~\ref{fig:topo}. 
The other terms proportional to products of the metric originate from the loops involving only two internal propagators; these terms are analytic in two of the momenta and give rise to contact terms when Fourier transformed back to position space. A discussion of the nature of these terms appears in Section \ref{sec:matching}. These contact terms drop out of the final polarization contractions that are compared to the results from holography. 

\section{Holographic computation of 3-point correlators}
\label{sec:holo3}

In this section we will compute the three-point correlators discussed
above, from the bulk theory, using AdS/CFT. 

We will work with the Poincare patch of AdS:
\be
\label{metricpoincare}
ds^2 = { dz^2 + \eta^{i j} d x_i d x_j \over z^2} ,
\ee
where $i,j$ run over the three boundary directions and we have set the AdS radius to $1$. So, all dimensionful quantities that follow are measured in these units.

\begin{figure}[h]
\begin{center}
 \includegraphics[width=2.5in]{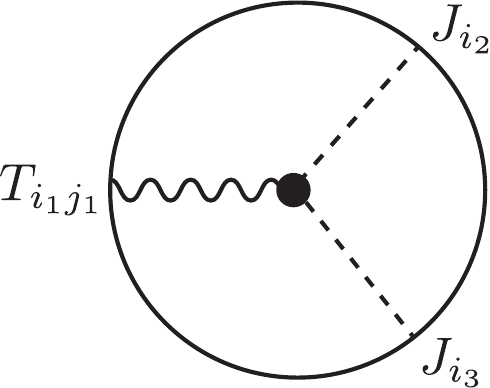}
 \caption{Witten diagram illustrating the holographic computation. The disk represent AdS$_4$, and the CFT is on its boundary. The holographic co-ordinate, $z$,
 is the radial direction. The wavy line is a bulk graviton $h_{\mu\nu}$, and the dashed line is the gauge field $A_\mu$.}
\label{fig:witten}
\end{center}
\end{figure}

The computation of the correlator requires us to evaluate the bulk
action to non-linear order, in the presence of certain solutions to
the linearized equations of motion. 
This corresponds to evaluating the ``Witten diagram'' in Fig.~\ref{fig:witten} which requires a three-point bulk interaction between
the gauge fields and the fluctuations of the metric. 

\subsection{Evaluation of the Bulk Action}

The first step in our computation is to write down the non-linear
 three-point interaction terms in the action. We can simplify our calculation by realizing that we 
are only interested in evaluating this action ``on-shell,'' (when the
gauge field and metric perturbation satisfy linearized equations of motion) and so
there are various terms that we can drop, as we will do below. 

The relevant part of the action is:
\be
S = 
{1 \over g_4^2} 
\int d^4 x \sqrt{-g} \left[-{1 \over 4} F_{\mu \nu} F_{\rho \sigma} g^{\mu \rho} g^{\nu \sigma} + \gamma C_{\mu \nu \rho \sigma} F_{\alpha \beta} F_{\gamma \delta} g^{\mu \alpha} g^{\nu \beta} g^{\rho \gamma} g^{\sigma \delta} \right].
\ee

First, we need to expand the Weyl tensor term in terms of the metric perturbation. We will use the conformal transformation properties of the Weyl tensor to write:
\be
\label{weyltransform}
C_{\alpha \beta \gamma \delta}\left({\eta_{\mu \nu} \over z^2} + h_{\mu \nu} \right) = {1 \over z^2} C_{\alpha \beta \gamma \delta}\left( \eta_{\mu \nu}  + z^2 h_{\mu \nu} \right) ,
\ee
where the Weyl tensor is written as a function of the metric.
For convenience, we define
\be
\widetilde{h}_{\mu \nu} = z^2 h_{\mu \nu}.
\ee
In what follows below, we will use the notation that:
\be
\begin{split}
C_{\alpha \beta \gamma \delta} &\equiv C_{\alpha \beta \gamma \delta}\left({\eta_{\mu \nu} \over z^2} + h_{\mu \nu} \right), \\
\widetilde{C}_{\alpha \beta \gamma \delta} &\equiv C_{\alpha \beta \gamma \delta}\left(\eta_{\mu \nu}  + \widetilde{h}_{\mu \nu} \right) ,
\end{split}
\ee
with similar conventions for other quantities like the Riemann and Ricci tensors. (A tilde comes
on top of quantities evaluated in the  flat space background metric, with the perturbation $\widetilde{h}$.)

We can choose a gauge --- both in flat space, and in AdS --- where the
metric fluctuation obeys:
\be
\label{condmet1}
\widetilde{h}^{z \mu} = 0.
\ee
It is easy to check that solutions to the equations of motion must be 
transverse and traceless:
\be
\label{condmet2}
\widetilde{h}_{\mu \nu} \eta^{\mu \nu} = 0 = \partial_{\rho} h_{\mu \nu}  \eta^{\mu \rho}.
\ee

If we know that  we will only have to evaluate the interaction vertex on wave-functions that obey \eqref{condmet1} and  \eqref{condmet2}, we can simplify the expressions for the Riemann tensor, the Ricci tensor, and the Ricci scalar in the linearized theory:
\be
\begin{split}
\widetilde{R}_{\alpha \mu \beta \nu} &= {1 \over 2} \left( \widetilde{h}_{\alpha \nu, \mu \beta} + \widetilde{h}_{\mu \beta, \nu \alpha} - \widetilde{h}_{\mu \nu, \alpha \beta} - \widetilde{h}_{\alpha \beta, \mu \nu} \right), \\
\widetilde{R}_{\alpha \beta} &= -{1 \over 2} \eta^{\mu \nu} \partial_{\mu} \partial_{\nu} \widetilde{h}_{\alpha \beta},\\
\widetilde{R} &= 0.
\end{split}
\ee
From this, we can obtain the Weyl tensor, which is: (here $d$ is the {\em boundary dimension}, and so $d + 1$ is the bulk dimension)
\be
\begin{split}
\widetilde{C}_{\alpha \mu \beta \nu} &= \widetilde{R}_{\alpha \mu \beta \nu} - {2 \over d - 1} \left(\eta_{\alpha [\beta} \widetilde{R}_{\nu] \mu}   - \eta_{\mu [\beta } \widetilde{R}_{\nu] \alpha} \right) + {2 \over d (d - 1)} \widetilde{R} \eta_{\alpha [\beta } \eta_{\nu] \mu} \\
&=  {1 \over 2} \left( \widetilde{h}_{\alpha \nu, \mu \beta} + \widetilde{h}_{\mu \beta, \nu \alpha} - \widetilde{h}_{\mu \nu, \alpha \beta} - \widetilde{h}_{\alpha \beta, \mu \nu} + \Box \left\{\eta_{\alpha [\beta} \widetilde{h}_{\nu] \mu} - \eta_{\mu [ \beta} \widetilde{h}_{\nu] \alpha} \right\}\right),
\end{split}
\ee
where, we have defined $\Box \equiv \eta^{\mu \nu} \partial_{\mu} \partial_{\nu}$ and used $d = 3$. 

However, this expression can be simplified considerably. With the understanding that $i,j,k,l$ run over the {\em boundary directions} and with $z$ representing the radial-direction, we need the following components of the Weyl tensor: 
\be
\label{weylcomponents}
\begin{split}
2 \widetilde{C}_{z i z j} &= {1 \over 2} \left[\sum_l \partial_l^2 - \partial_z^2 \right] \widetilde{h}_{i j}, \\
2 \widetilde{C}_{z i j k} &= \partial_{k} \partial_{z}\widetilde{h}_{i j}  -\partial_j \partial_z \widetilde{h}_{i k}, \\
2 \widetilde{C}_{i j k l} &=  {1 \over 2} \left[\partial_z^2  - \sum_l \partial_l^2 \right] \left(\eta_{i k} \widetilde{h}_{j l} - \eta_{i l} \widetilde{h}_{k j} - \eta_{j k} \widetilde{h}_{i l} + \eta_{j l} \widetilde{h}_{i k} \right).
\end{split}
\ee
In evaluating the first two lines, we have used the conditions \eqref{condmet1} and \eqref{condmet2}. In evaluating the last line, we have used the fact that the Weyl tensor vanishes identically in $3$-dimensions. This might suggest that only the additional term involving the $z$-derivatives survives; however, one needs to be careful about the factor in front of the Laplacian, which is dimension dependent. When we take all of this into account, we get the expression above.\footnote{This almost--- but not quite --- agrees with the results of Ref.~\onlinecite{Maldacena:2011nz}.  In particular the first line of
  \eqref{weylcomponents} does not agree with the first line of (2.12)
  of Ref.~\onlinecite{Maldacena:2011nz} in general, and neither does the last
  line.  However, the expressions  do agree if we are evaluating this tensor on
  a solution of the form $h_{i j} = \ep_{i j} e^{-\norm{k} z + i k
    \cdot x}$, which was the case under consideration in that paper.}

With these results for the flat-space Weyl tensor, the expression for the Weyl tensor in AdS is also fixed by the relation \eqref{weyltransform}.  We should point out that while we have not written all the non-zero components above, the components that we have written, and the symmetries of the Weyl tensor fix everything.

To evaluate the interaction vertex, we also note the fact that, for the evaluation of the 
three point function under consideration, the non-Abelian terms in the field-strength are unimportant. So, in what follows below, we simply take:
\be
F_{\mu \nu} = \partial_{\mu} A_{\nu} - \partial_{\nu} A_{\mu},
\ee
and choose a gauge where
\be
\label{condym}
A_{z} = 0, \quad \partial_i A^i = 0.
\ee

To finally evaluate the interaction vertex in AdS, we will use
the explicit forms of the wave functions for the gauge field and 
for the graviton. These are given by:
\be
\label{wavefuncs}
\begin{split}
A_i &= \ep_i e^{- \norm{k} z} e^{i \vect{k} \cdot \vect{x}}, \\
h_{i j} &= {1 \over z^2} e_{i j} e^{-\norm{k} z} \left(1 + \norm{k} z\right) e^{i \vect{k} \cdot \vect{x}},\\
\widetilde{h}_{i j} &= e_{i j} e^{-\norm{k} z} \left(1 + \norm{k} z\right) e^{i \vect{k} \cdot \vect{x}}.
\end{split}
\ee
See Ref.\onlinecite{Raju:2012zs} for further details on the notation. Below, we will use $R_m$ to denote the radial part of the wave function
of the $m^{\text{th}}$ particle:
\be
R_1 = (1 + \norm{k_1} z)e^{-\norm{k_1} z}, \quad R_2 = e^{-\norm{k_2} z}, \quad R_3 = e^{- \norm{k_3} z},
\ee
and also use the notation: $\dot{f} \equiv {\partial f \over \partial z}$.

We now need to evaluate the variation of the action 
 to first order in the metric perturbation $h$, and second order in the gauge field. This is appropriate, since we wish to compute a three point function involving one 
stress tensor and two currents. Since the Weyl tensor vanishes in pure AdS, and we have no gauge field background either, the variation in the Weyl-gauge term is simply its value in the presence of the perturbation, 
\be
{
g_4^2 
 \over \gamma} \delta S_1 = \int d z \sqrt{-g} C_{\mu \nu \rho \sigma}  F_{\alpha \beta} F_{\gamma \delta} g^{\mu \alpha} g^{\nu \beta} g^{\rho \gamma} g^{\sigma \delta}  ;
\ee
here and below we drop the integral over the flat three-dimensional spacetime whose role is to ensure conservation of momentum
with $\vect{k}_1 + \vect{k}_2 + \vect{k}_3 = 0$.

Let us now evaluate the different contractions that appear in the expression above, keeping track of the numerical factors. 
\begin{itemize}
\item
First of all, we note that given the expressions for the wave functions in \eqref{wavefuncs} above, we can always replace a derivative $\partial_j \rightarrow i k_j$. Each term in the contraction has two such spatial derivatives leading to an overall minus sign. 
\item
Secondly, the Weyl tensor is anti-symmetric under the interchange of the first two or the third and fourth indices. Since the field strength is also anti-symmetric, we get a factor of $4$ by summing over these permutations. 
\item
Finally, there is a factor of ${1 \over 2}$ in \eqref{weylcomponents}, but we have to keep in mind that we need to sum over the two possible permutations of the gauge-fields in the Witten diagram. 
\end{itemize}

Therefore, we have:
\be
\begin{split}
 \delta S_{11} &= C_{z i z j}  F^{z i} F^{z j} +  C_{i z j z}  F^{i z} F^{j z} +  C_{i z z j}  F^{i z} F^{z j} +  C_{z i j z}  F^{z i} F^{j z } \\ &= -2 z^6 (\vect{\ep_1} \cdot \vect{\ep_2}) (\vect{\ep_1} \cdot \vect{\ep_3}) \left( \norm{k_1}^2 R_1 + \ddot{R_1} \right) \dot{R_2} \dot{R_3},\\
\delta S_{12} &=  C_{z i j k}  F^{z i}  F^{j k} +  C_{i z j k}  F^{i z}  F^{j k} +  C_{i j z k}  F^{i j}  F^{z k} +  C_{i j k z}  F^{i j}  F^{k z} \\ &=  -2 z^6 \Big[ \left\{(\vect{k_1} \cdot \vect{\ep_3}) (\vect{k_3} \cdot \vect{\ep_1}) - (\vect{k_1} \cdot \vect{k_3}) (\vect{\ep_1} \cdot \vect{\ep_3}) \right\} (\vect{\ep_1} \cdot \vect{\ep_2}) \dot{R_1} \dot{R_2} R_3 \\ &+  \left\{(\vect{k_1} \cdot \vect{\ep_2}) (\vect{k_2} \cdot \vect{\ep_1})- (\vect{k_1} \cdot \vect{k_2}) (\vect{\ep_1} \cdot \vect{\ep_2}) \right\} (\vect{\ep_1} \cdot \vect{\ep_3}) \dot{R_1} R_2  \dot{R_3} \Big],\\
\delta S_{13} &=  C_{i j k l}  F^{i j}  F^{k l} \\ &= -4 z^6 \Big[(\vect{k_2} \cdot \vect{k_3}) (\vect{\ep_3} \cdot \vect{\ep_1}) (\vect{\ep_2} \cdot \vect{\ep_1}) - (\vect{k_2} \cdot \vect{\ep_1}) (\vect{k_3} \cdot \vect{\ep_2}) (\vect{\ep_1} \cdot \vect{\ep_3})  \\ 
&\hphantom{ -2 \Big[}
- (\vect{k_3} \cdot \vect{\ep_1}) (\vect{k_2} \cdot \vect{\ep_3}) (\vect{\ep_2} \cdot \vect{\ep_1}) + (\vect{k_2} \cdot \vect{\ep_1}) (\vect{k_3} \cdot \vect{\ep_1}) (\vect{\ep_2} \cdot \vect{\ep_3}) \Big]  \left(\ddot{R_1} + \norm{k_1}^2 R_1 \right) R_2 R_3. \\
\end{split}
\ee
Let us make a comment about the overall power of $z$. We get four-factors of $z^2$ from the four inverse metric components that are required to raise the indices of $F$. However, we get one factor of ${1 \over z^2}$ from $C$. This is what leads to the overall $z^6$ outside. Also, we caution the reader that when we write $\vect{\ep_1} \cdot \vect{\ep_2}$ above, and other such expressions involving the dot-product of three-dimensional vectors, this dot-product is taken with the flat space metric:
\be
\vect{\ep_1} \cdot \vect{\ep_2} \equiv \ep_{1 i} \ep_{2 j} \eta^{i j}.
\ee

The variation of the full Weyl-gauge term in the action is just the sum of the terms above:
\be
\label{var1}
\begin{split}
{
g_4^2 
 \over \gamma} \delta S_1 = \int d z \sqrt{-g} \left[\delta S_{11} + \delta S_{12} + \delta S_{13} \right].
\end{split}
\ee

There is, of course, another term that contributes to the three-point function,
which comes from the interaction of the metric perturbation with the 
stress tensor of the gauge field. This evaluates to:
\be
\label{var2}
g_4^2 
\delta S_2 = \int \sqrt{-g} d z \left[{1 \over 2} F_{\mu \nu} F_{\rho \sigma} \eta^{\mu \alpha}  h_{\alpha \beta} \eta^{\beta \rho} \eta^{\nu \sigma}\right] z^6.
\ee
Note that the conditions \eqref{condmet1}, \eqref{condmet2} mean we can drop the term that comes from the variation of $\sqrt{-g}$. We also have an overall minus sign because $\delta g^{\mu \nu} = -g^{\mu \rho} h_{\rho \sigma} \delta g^{\sigma \nu}$. The overall factor of $z^6$ comes from the four inverse metric factors, but it is important to remember that one needs to include the ${1 \over z^2}$ in $h_{\alpha \beta}$ from \eqref{wavefuncs}.

We can write
\be
\begin{split}
g_4^2 
\delta S_2 = -\int  d z  \Bigg\{&\Big[(\vect{k_2} \cdot \vect{\ep_1}) (\vect{k_3} \cdot \vect{\ep_1}) (\vect{\ep_2} \cdot \vect{\ep_3}) - (\vect{k_2} \cdot \vect{\ep_1}) (\vect{\ep_3} \cdot \vect{\ep_1}) (\vect{\ep_2} \cdot \vect{k_3}) \\ &- (\vect{k_3} \cdot \vect{\ep_1}) (\vect{\ep_2} \cdot \vect{\ep_1}) (\vect{\ep_3} \cdot \vect{k_2}) 
+ (\vect{k_2} \cdot \vect{k_3}) (\vect{\ep_2} \cdot \vect{\ep_1}) (\vect{\ep_3} \cdot \vect{\ep_1}) \Big] R_1 R_2 R_3 \\
&- (\vect{\ep_1} \cdot \vect{\ep_2}) (\vect{\ep_1} \cdot \vect{\ep_3}) R_1 \dot{R_2} \dot{R_3} \Bigg\}.
\end{split}
\ee
Note that we have regained a minus sign from the two factors of $i$ that get pulled down in the differentiation, although this does not occur in the last term above where we have a $z$-derivative instead. Also note that all factors of $z$ are gone, when we account for the $\sqrt{-g}$ and the factor of ${1 \over z^2}$ in $h_{\alpha \beta}$ from \eqref{wavefuncs}. 

As a final step in evaluating the 3-point function we now  need to do the radial integrals in \eqref{var1} and \eqref{var2}.  First, let us  do the radial integrals in \eqref{var1}. Note that once we account for the fact that $\sqrt{-g} = {1 \over z^4}$, there is an overall factor of $z^2$ in every radial integral.
These are
\begin{align}
&\int z^2 \ddot{R_1} R_2 R_3 \, dz = \frac{2 \norm{k_1}^2 (2
   \norm{k_1}-\norm{k_2}-\norm{k_3})}{(\norm{k_1}+\norm{k_2}+\norm{k_3})^4}, \\
&\int z^2 \dot{R_1} \dot{R_2} R_3 \, dz = \frac{6 \norm{k_1}^2 \norm{k_2}}{(\norm{k_1}+\norm{k_2}+\norm{k_3})^4}, \\
&\int z^2 \dot{R_1} \dot{R_2} R_3 \, dz = \frac{6 \norm{k_1}^2 \norm{k_3}}{(\norm{k_1}+\norm{k_2}+\norm{k_3})^4}, \\
&\int z^2 R_1 \dot{R_2} \dot{R_3} \, d z = \frac{2 \norm{k_2} \norm{k_3} (4
   \norm{k_1}+\norm{k_2}+\norm{k_3})}{(\norm{k_1}+\norm{k_2}+\norm{k_3})^4}.\\ 
\end{align}
Now, we turn to the radial integrals in \eqref{var2}. These are
\begin{align}
&\int R_1 R_2 R_3 \, d z=  \frac{2
   \norm{k_1}+\norm{k_2}+\norm{k_3}}{(\norm{k_1}+\norm{k_2}+\norm{k_3})^2}, \\
&\int R_1 \dot{R_2} \dot{R_3} \, dz= \norm{k_2} \norm{k_3} \frac{2 \norm{k_1}+\norm{k_2}+\norm{k_3}}{(\norm{k_1}+\norm{k_2}+\norm{k_3})^2}. \\
\end{align}
All these integrals are convergent.

\subsection{Final Bulk Answers in the Spinor Helicity Formalism}
The expressions for the bulk action, and the radial integrals above in
principle give us all the information we need about the boundary
correlator. However, to extract some physics from this, it is convenient
to choose  various ``helicities'' for the 
stress-tensor and the currents and then write down the answer in the
spinor-helicity formalism that was outlined above. 

We only need to consider the following three choices of helicities:
\begin{enumerate}
\item
Both currents, and the stress tensor have negative helicity.
\item
The stress tensor and one current has negative helicity, but the other current has positive helicity.
\item
The stress tensor has positive helicity, and the two currents have negative helicity.
\end{enumerate}
All other possibilities can be obtained from these ones by permuting the
two currents and/or using parity. 

The use of the spinor-helicity formalism considerably simplifies the
 algebraic expressions involved in the answers.  The reader who is
 interested in the algebra that enters this simplification should
 consult the accompanying Mathematica file \cite{simplifybulk}. Here, we simply
 present the final answers.
    
For the case where all the helicities are negative, we have the following expression.
\be
\label{adsmmm}
\mathcal{K}_{\text{ads}}^{---}(\vect{k_1}, \vect{k_2}, \vect{k_3}) = -24 \gamma {\dotl[\la_2,\la_1]^2 \dotl[\la_3,\la_1]^2 \norm{k_1} \over 
g_4^2  
E^4},
\ee
where we have defined:
\be
E \equiv \norm{k_1} + \norm{k_2} + \norm{k_3}.
\ee
It is natural for this expression (considered as an analytic function
of $E$) to have a pole at $E = 0$, and in fact the residue at this
pole is related to the four dimensional flat space amplitude of a graviton and two
gluons as pointed out in Ref.~\onlinecite{Raju:2012zr}. 
We also note that the usual gravitational interaction does not contribute to this helicity combination at all, and the entire combination comes from the Weyl interaction.

When the stress tensor and the first current insertion are dotted with
negative helicity polarization vectors 
and the second current is dotted with a positive helicity polarization
vector, we find:
\be
\label{adsmmp}
\mathcal{K}^{--+}_{\text{ads}} =  -{\dotl[\la_2,\la_1]^4 \left(\norm{k_2} + \norm{k_3} - \norm{k_1} \right)^2 \left(2 \norm{k_1} + \norm{k_2} + \norm{k_3} \right)\over  2 \dotl[\la_3,\la_2]^2 \norm{k_1}^2  
g_4^2 
E^2 }.
\ee
In this case, we find that the Weyl interaction does not contribute to this helicity combination, whereas the usual gravitational interaction does.

Finally, we come to the the case where the stress-tensor has positive helicity and the 
two currents have negative helicity. For this correlator, we have:
\be
\label{adspmm}
\mathcal{K}^{+--}_{\text{ads}} = 0.
\ee
Neither the Weyl nor the gravitational interaction contribute to this helicity combination!

It is useful to check that these answers indeed have the expected behaviour
under scaling. Recall that the stress tensor has dimension $3$, and the
two conserved currents have dimension $2$ each. Fourier transforming
the 3-position variables gives us a dimension of $-9$, of which the
momentum space $\delta$-function that we have suppressed above soaks
up $-3$. So, we expect the net dimension in momentum space to be $1$,
which is true of all the expressions above. 

The spinor helicity formalism only makes the Lorentz group on the
boundary manifest. It is possible to check that these answers also
satisfy the constraints of special conformal transformations as
indicated in Ref.~\onlinecite{Maldacena:2011nz}, but this
is a slightly more involved calculation. 

\section{Matching the Answers \label{sec:matching}}

In this section, we will match the answers of the CFT computations of Sec.~\ref{sec:CFT3} (and Appendix~\ref{app:two_point}) 
with the AdS answers of Sec.~\ref{sec:holo3}. This will allow us to determine the values of
physical parameters in the bulk, that would reproduce the free
answers.
\subsection{Scalars} 
\paragraph{$--+$ Helicity}
Let us start with \eqref{mmpfree}, which we can write as:
\be
\begin{split}
{1 \over N_s} \mathcal{K}^{--+}_{s} &= \frac{\dotl[\la_2, \la_1]^4 (-\norm{k_{1}}+\norm{k_{2}}+\norm{k_{3}})^2  
} {32
   \dotl[\la_3, \la_2]^2  \norm{k_{1}}^2 \norm{k_{2}} \norm{k_{3}}  (\norm{k_{1}}+\norm{k_{2}}+\norm{k_{3}})^2} 
\\ &\times \left[(\norm{k_1} + \norm{k_2} + \norm{k_3})^2 (\norm{k_2} + \norm{k_3}) - 2 \left(2 \norm{k_1} + \norm{k_2} + \norm{k_3} \right) \norm{k_2} \norm{k_3} \right] \\
&= {1 \over N_s} \widetilde{\mathcal{K}}^{--+}_{s} + {\cal C}^{--+}_s.
\end{split}
\ee
Here, 
\be
\widetilde{\mathcal{K}}^{--+}_{s} =  - N_s {\dotl[\la_2,\la_1]^4 \over   \dotl[\la_3,\la_2]^2}\frac{(-\norm{k_1}+\norm{k_2}+\norm{k_3})^2 (2 \norm{k_1}+\norm{k_2}+\norm{k_3})}{16 \norm{k_1}^2
   (\norm{k_1}+\norm{k_2}+\norm{k_3})^2},
\ee
 has exactly the same functional form as the answer obtained from the AdS calculation in \eqref{adsmmp} and we have defined:
\be
\label{cmmp}
{\cal C}_{s}^{--+} =  \frac{\dotl[\la_2, \la_1]^4 (-\norm{k_{1}}+\norm{k_{2}}+\norm{k_{3}})^2 \left(\norm{k_2} + \norm{k_3} \right) 
} {32
   \dotl[\la_3, \la_2]^2  \norm{k_{1}}^2 \norm{k_{2}} \norm{k_{3}} } .
\ee
We will now show that \eqref{cmmp} is purely a contact term. 
In fact we can write
\be
\label{cmmprewritten}
{\cal C}_{s}^{--+} = {1 \over 8} \left(\vect{\ep_1} \cdot \vect{\ep_2}\right) \left(\vect{\ep_1} \cdot \vect{\ep_3} \right) \left(\norm{k_2} + \norm{k_3}\right) .
\ee
To check the equivalence of \eqref{cmmprewritten} and \eqref{cmmp}, we note that:
\be
\begin{split}
&{-1 \over 8} \left(\vect{\ep_1} \cdot \vect{\ep_2}\right) \left(\vect{\ep_1} \cdot \vect{\ep_3} \right) \left(\norm{k_2} + \norm{k_3}\right) \\
&= {-1 \over 32 \norm{k_1}^2 \norm{k_2} \norm{k_3}} \dotl[\la_1, \la_2]^2 \dotlm[\la_1, \lb_3]^2 \left(\norm{k_2} + \norm{k_3} \right)\\
&= {1 \over 32 \norm{k_1}^2 \norm{k_2} \norm{k_3}} {\dotl[\la_1, \la_2]^4 \dotl[\lb_2, \lb_3]^2 \over E^2}  \left(\norm{k_2} + \norm{k_3} \right) \\
&={1 \over 32 \norm{k_1}^2 \norm{k_2} \norm{k_3}} {\dotl[\la_1, \la_2]^4 \over \dotl[\la_2, \la_3]^2} \left(\norm{k_2} + \norm{k_3} - \norm{k_1} \right)^2  \left(\norm{k_2} + \norm{k_3} \right),
\end{split}
\ee
where the last line is manifestly the same as \eqref{cmmp}.

However, we can write \eqref{cmmprewritten} as 
\be
{\cal C}_{s}^{--+} = {1 \over 32} e_{1 i_1 j_1} \ep_{2 i_2} \ep_{3 i_3} \left[\eta^{i_1 i_2} \eta^{j_1 i_3} \left( \norm{k_2} + \norm{k_3} \right) \right].
\ee
The term in the square brackets is the ``bare'' correlator, before contracting with the polarization vectors, and this is the term we should Fourier transform to position space. In position space, this is evidently a contact term. 

The general rule is that a term that is ``analytic'' in two of the momenta yields a contact term when Fourier transformed to position space. In this case, we notice, for example, that after adding the overall momentum conserving delta function we have:
\be
\int \norm{k_2}  \delta(\vect{k_2} + \vect{k_3} + \vect{k_1}) e^{i \sum \vect{k_m} \cdot \vect{x_m}} \prod d^3 \vect{k_m}= (2 \pi)^3 \delta(x_1 - x_3) \int \norm{k_2} e^{i \vect{k_2} \cdot (\vect{x_2} - \vect{x_3})} d^3 \vect{k_2}.
\ee

Contact terms in correlators are very subtle since they depend on the
precise definition of the correlator, and also on the regulator used
to compute it. While they might have physical significance under some
circumstances, in this paper, we will just drop these additional
$\delta$ function terms and work with $\tilde{{\cal
    K}}_s^{--+}$ instead of  ${{\cal
    K}}_s^{--+}$. 

\paragraph{$+--$ Helicity}
It turns out that the free answer \eqref{freepmm} is entirely a contact term in this case! We note that 
\be
\begin{split}
&{1 \over 4} (\vect{\ep_1} \cdot \vect{k_2}) (\vect{\ep_1} \cdot \vect{k_3}) (\vect{\ep_2} \cdot \vect{\ep_3}) \left(\norm{k_2} + \norm{k_3} \right)\\ 
&= {1 \over 32 \norm{k_1}^2 \norm{k_2} \norm{k_3}} \dotl[\la_2, \la_3]^2 \dotl[\lb_1, \lb_2] \dotl[\lb_1, \lb_3] \dotlm[\lb_1, \la_2] \dotlm[\lb_1, \la_3] \left(\norm{k_2} + \norm{k_3} \right) \\
&= {1 \over 32} \dotl[\la_2, \la_3]^4 (\norm{k_1} + \norm{k_3} - \norm{k_2})^2 (\norm{k_1} + \norm{k_2} - \norm{k_3})^2 \left(\norm{k_2} + \norm{k_3} \right) \\
&= {1 \over N_s} \mathcal{K}_{s}^{+--}.
\end{split}
\ee
This is consistent with the fact that both the Weyl interaction and the ordinary gravitational interaction yield $0$ in the AdS calculation \eqref{adspmm}. For notational consistency, we can set:
\be
\widetilde{\mathcal{K}}^{+--}_{s} = 0, \quad  {\cal C}_{s}^{+--} = {1 \over N_s} \mathcal{K}_{s}^{+--}.
\ee

\paragraph{$---$ Helicity}
Turning finally to \eqref{freemmm}, we see that this expression can be written as:
\be
{1 \over N_s} \mathcal{K}^{---}_{s} = {1 \over N_s} \widetilde{\mathcal{K}}^{---}_{s} + {\cal C}_{s}^{---},
\ee
where 
\be
\widetilde{\mathcal{K}}^{---}_{s} =  N_s \frac{\dotl[\la_2, \la_1]^2 \dotl[\la_3, \la_1]^2}{4}{ \left(\frac{ \norm{k_{1}}}{(\norm{k_{1}}+\norm{k_{2}}+\norm{k_{3}})^4} \right)},
\ee
 has exactly the same functional form as the AdS answer \eqref{adsmmm} and the contact term ${\cal C}_{s}^{---}$ is:
\be
{\cal C}_{s}^{---} = {-1 \over 8} (\vect{\ep_1} \cdot \vect{\ep_2}) (\vect{\ep_1} \cdot \vect{\ep_3}) \left(\norm{k_2} + \norm{k_3} \right).
\ee

\subsubsection{Value of $\gamma$}
Our final task is to find the value of $\gamma$ for free-scalars. To make the normalization of the two-point functions drop out, we can simply consider the ratio:
\be
{\widetilde{\mathcal{K}}^{---}_{s} \over \widetilde{\mathcal{K}}^{--+}_{s}} = -12 \gamma_s {\mathcal{K}^{---}_{\text{ads}} \over \mathcal{K}^{--+}_{\text{ads}}}.
\ee
Since the two ratios above should be equal, we find that we should set:
\be
\gamma_{s} = -{1 \over 12},
\label{gammas}
\ee
where we have added a subscript to distinguish it from the value for free-fermions that we will find below.

We end by  pointing out a  very interesting feature of the answers \eqref{adsmmm},
\eqref{adspmm} and \eqref{adsmmp}: there is no term where the ordinary
interaction and the Weyl interaction contribute simultaneously. If we
had a term where the two interactions contributed simultaneously, we
could have fixed $\gamma$ by looking at the functional form of the
answer. However, $\gamma$ appears as a simple ratio of two
answers and so one needs to be extremely careful in determining all the signs and numerical prefactors in the expressions for the various 3-point functions correctly. 

\subsubsection{Value of $G_N$}
We can also set the value of $G_N$ from our calculations. Although $G_N$ does not appear in the three-point computations above, it does appear in the computation of the two-point function for the stress-tensor from the bulk using the action \eqref{heaction}. If we write the results for the two point functions in Appendix \ref{app:two_point} and Appendix \ref{twopointstressgrav} as:
\be
\begin{split}
&\ep_{1,i_1} \ep_{1,i_2} \ep_{2,i_3} \ep_{2,i_4} \langle T_s^{i_1 i_2}(\vect{k}) T_s^{i_3 i_4}(-\vect{k}) \rangle = C_{T,s} \norm{k}^3 (\vect{\ep_1} \cdot \vect{\ep_2})^2, \\
&\ep_{1,i_1} \ep_{2,i_2} \langle J_s^{i_1} J_s^{i_2} \rangle = -C_{J,s} \norm{k}  (\vect{\ep_1} \cdot \vect{\ep_2}),
\end{split}
\ee
then we should demand that the normalization independent quantities be equal:
\be
\label{normalindependent}
\begin{split}
{1 \over \sqrt{C_{T,s}} C_{J,s}} \widetilde{\mathcal{K}}^{---}_{s} = {1 \over \sqrt{C_{T,\text{ads}}} C_{J,\text{ads}}} \mathcal{K}^{---}_{\text{ads}},\\
{1 \over \sqrt{C_{T,s}} C_{J,s}} \widetilde{\mathcal{K}}^{--+}_{s} = {1 \over \sqrt{C_{T,\text{ads}}} C_{J,\text{ads}}} \mathcal{K}^{--+}_{\text{ads}}.
\end{split}
\ee
We have, from the results for two point functions:
\be
\begin{split}
&C_{T,s} = {N_s \over 256}, \quad C_{J,s} = {N_s \over 16},\\
&C_{T,\text{ads}} = {1 \over \pi G_{N,s}}, \quad C_{J,\text{ads}} = {1 \over g_{4,s}^2}.
\end{split}
\ee
This leads to the scalar contribution 
\be
\frac{1}{G_{N,s}} = {\pi N_s \over 1024 L^2},
\ee
where we have reinstated the dimensionful factor of the AdS radius.

Note that, with this choice, the quantities $C_{T,s}$ and $C_{T, \text{ads}}$ do not agree and this is a sign of the fact that, with our conventions, the stress-tensor of the bulk theory is normalized differently from that of the boundary theory. This, in turn, results from our choice of $Z$ above \eqref{gravbulkbound}. This choice was made to yield a particularly simple graviton bulk to boundary propagator, and to get $C_{T,s}$ to match with $C_{T,\text{ads}}$ we should have chosen $Z = {-d \over 4 \pi G_N}$, which is twice the choice that we have made currently.

\subsubsection{Value of $g_4^2$}
Note that $g_{4,s}^2$ does not appear in the quantities \eqref{normalindependent} at all since it cancels between the three-point and the two-point functions. However, we can choose a value by demanding that the two-point functions of the currents be equal in the bulk and the boundary. Imposing:
\be
C_{J,\text{ads}} = C_{J,s},
\ee
we can set:
\be
g_{4,s}^2 = {16 \over N_s}.
\ee

\subsection{Fermions}

The analysis for the fermionic answers is almost identical, so we will not repeat it in detail here. However, with a little work, (see the Mathematica file \cite{simplifyfermions}) we find that we can write:
\be
\begin{split}
&{1 \over N_f} \mathcal{K}^{---}_{f} = {1 \over N_f} \widetilde{\mathcal{K}}^{---}_{f} + {\cal C}_f^{---}, \\
&{1 \over N_f} \mathcal{K}^{--+}_{f} = {1 \over N_f} \widetilde{\mathcal{K}}^{--+}_{f} + {\cal C}_f^{--+}, \\
&{1 \over N_f} \mathcal{K}^{+--}_{f} = {1 \over N_f} \widetilde{\mathcal{K}}^{+--}_{f} + {\cal C}_f^{+--}, \\
\end{split}
\ee
where
\be
\widetilde{\mathcal{K}}^{---}_{f} =  - \widetilde{\mathcal{K}}^{---}_{s}, \quad \widetilde{\mathcal{K}}^{--+}_{f} =   \widetilde{\mathcal{K}}^{--+}_{s}, \quad \widetilde{\mathcal{K}}^{+--}_{f} =  \widetilde{\mathcal{K}}^{+--}_{s} = 0,
\ee
and the analytic remainders are:
\be
{\cal C}_f^{---} = {1 \over 2} {\cal C}_s^{---}, \quad  {\cal C}_f^{--+} =  {1 \over 2} {\cal C}_s^{--+} , \quad {\cal C}_f^{+--} = {1 \over 2} {\cal C}_s^{+--},
\ee
which are half those of the scalar-case above.

Thus, we immediately see that for free-fermions, we have 
\be
\gamma_{f} = -\gamma_{s}= {1 \over 12}.
\label{gammaf}
\ee
A standard computation of the fermion 2-point functions shows that as for the scalars, we now have
\bea
g_{4,f}^2 &=& {16 \over N_f}, \nn
\frac{1}{G_{N,f}} &=& {\pi N_f \over 1024 L^2}.
\eea
Of course, the CFT only has a single $G_N$ which is simply $1/G_N = 1/G_{N,f} + 1/G_{N,s}$ at this order in $1/N_F$.

\subsection{Topological Current} 
To obtain the value of $\gamma$ for the topological current, we do not
need to do any additional work. The analysis for the topological
current proceeds in the following sequence of steps:
\begin{enumerate}
\item
First, we can ignore the third line of \eqref{ktans} which includes
terms like $\eta^{v2 j1}, \eta^{v2 i1}$ etc. since they are
analytic in two of the momenta. This leaves us with the terms
involving ${\cal K}_s$ and ${\cal K}_f$.
\item
Instead of contracting ${\cal K}_f$ and ${\cal K}_s$ with the polarization vectors
$\vect{\ep_{2}}$ and $\vect{\ep_{3}}$, we instead need to contract them with the vectors: $\vect{\ep_2} \times \vect{k_2} /\norm{k_2}$ and $\vect{\ep_3} \times \vect{k_3}/\norm{k_3}$.\footnote{We need to be careful because we are in Lorentzian space, and the ordinary rules for the cross-product will take us from two vectors with lowered indices to a vector with a raised index.}  
\item
However, this returns the original polarization vectors, up to a sign
that depends on the helicity. In particular
\be
\vect{\ep_2} \times \vect{k_2}/\norm{k_2} = h_2 \vect{\ep_2},
\ee
where $h_2$ is the helicity of current $2$.
 A similar formula holds for current $3$.
\item
Therefore we get the same amplitudes as earlier up to a sign that is $1$ if both currents have the same helicity and $-1$ if the currents have opposite helicities.
\end{enumerate}
This chain of argument immediately yields
\be
\gamma_t =  \frac{(N_s - N_f)}{12 (N_s + N_f)}.
\ee

\section{Position space correlators and energy flux} \label{sec:ajay}

In this section, we provide an alternate route to fix the value of $\gamma$ using the 3-point functions $\langle TJJ \rangle$ in position space. In particular, we will extend the calculation of energy flux in Ref.~\onlinecite{Hofman:2008ar} to arbitrary spacetime dimensions $d$ and by comparing it with the holographic results, we relate $\gamma$ to the parameters in the 3-point correlator of a general CFT
obtained by Osborn and Petkou \cite{Osborn:1993cr,Erdmenger:1996yc}. The latter parameters are known for free CFTs, and so we will obtain
an alternate derivation of the $N_F \rightarrow \infty$ limits of $\gamma_s$ and $\gamma_f$, consistent with our previous results.

In a CFT, we consider a thought experiment in which a localized disturbance or state is created by the insertion of a conserved vector current $(\vect{\epsilon}\cdot\vect{J})$, where $\vect{\epsilon}$ is a fixed, spatial polarization vector. We assume that this local disturbance injects a fixed energy $E$ and the system evolves in time. Now, we can place calorimeters at large distances and further study the anisotropic distribution of energy. In this experiment, a particular quantity, that is the energy flux escaping to the null infinity, will take a very simple form. If the direction of the null infinity is given by the unit vector $\vect{n}$, the energy flux collected by the calorimeter will be given by:
\bea
\langle \mathcal{E}(\vect{n})\rangle&\,=\,&\frac{\langle0|(\vect{\epsilon}^*\cdot \vect{J}^\dagger)\mathcal{E}(\vect{n})(\vect{J}\cdot\vect{\epsilon})|0\rangle}{ \langle0|(\vect{\epsilon}^*\cdot \vect{J}^\dagger)(\vect{J}\cdot \vect{\epsilon})|0\rangle} \label{ed1} \\
&\,=\,&  \frac{E}{\Omega_d} \left[ 1 + \mathcal{A}\left(\frac{|\vect{\epsilon}\cdot\vect{n}|^2 }{ |\vect{\epsilon}|^2}-\frac{1}{d-1}\right) \right]  \,,
\label{ed2}
\eea
This form of the energy flux is completely fixed by the energy conservation and $O(d-1)$ symmetry of the construction. Here $\mathcal{E}(\vect{n})$ is the energy flux operator, to be introduced shortly in \eqref{ed3}. The total energy injected by the perturbation is $E$ and $\Omega_d=2\,\pi^{\frac{d-1}{2}}/\Gamma(\frac{d-1}{2})$ is the area of the unit $(d-2)$-sphere. Further, $\mathcal{A}$ is a constant which characterizes the CFT. As pointed out after Eq.~\eqref{et}, the three point function $\langle TJJ \rangle$ in real space is completely determined by $C_J$, $C_T$ and an additional constant. The coefficient $\mathcal{A}$ is related to this additional constant and in holography, it is related to the coupling constant $\gamma$ in \eqref{Smyers}. In this section, we will find $\mathcal{A}$ through field theory and holographic calculations, and by comparing the results we will fix $\gamma$ for free scalar and fermionic field theories. First, we will begin with the 3-point function $\langle TJJ \rangle$ in position space, which is specified by Osborn and Petkou \cite{Osborn:1993cr}, and calculate energy density \eqref{ed1} for CFTs.

\subsection{$\mathcal{A}$ in CFTs}

To set up the calculations on field theory side, we work with Minkowski metric with `mostly positive' signature. In our thought experiment, we place the calorimeter at large distance along $x^1$ direction and hence the unit vector $n^i=\delta^i_1$. To measure the energy along the null infinity, it is convenient to use the light-cone coordinates, which we define as $x^{\pm}=x^0\pm x^1$. Then, the energy flux operator is given by \cite{Hofman:2008ar, Buchel:2009sk}
\be
\mathcal{E}(\vect{x_1},\vect{n})\,=\,\int dx_1^-\left[\lim_{x_1^+ \to \infty} \left(\frac{x_1^+-x_1^-}{2}\right)^{d-2} T_{--}(x_1^+,x_1^-) \right]\,,
\label{ed3}
\ee
where $T_{--}$ is the component of the stress energy tensor. Now to fix $\mathcal{A}$, it is sufficient to calculate the energy one point function for a state created by the operator $(\vect{J}\cdot\vect{\epsilon})$, which appears in the numerator of \eqref{ed1}. So the calculation will boil down to using the expression for three point function $\langle J_i^\dagger(\vect{x_2})T_{--}(\vect{x_1},\vect{n})J_j(\vect{x_3})\rangle$ and performing various integrations. We can simplify these integrations by using symmetries of the construction. In the correlations $\langle J_i^\dagger(\vect{x_2})T_{--}(\vect{x_1},\vect{n})J_j(\vect{x_3})\rangle$, we can use translation invariance to set $\vect{x_3}=0$. By aligning the calorimeter along $n^i=\delta^i_1$, we have also fixed $\vect{x_1}=\{x_1^0,x_1^1,0,\dots\}$. With these simplifications, we will only need to integrate over the coordinates $\vect{x_2}=\vect{x}=\{x^0,x^1,x^2,\dots\}$. We further choose the spatial polarization vector $\vect{\epsilon}$ to be $\vect{\epsilon}=\{\epsilon^0,\epsilon^1,\epsilon^2,\epsilon^3\dots\}=\{0,\cos\theta,\sin\theta,0,\dots \}$. In this notation, we clearly have $|\vect{\epsilon}\cdot\vect{n}|=\cos\theta$ and the numerator of \eqref{ed1} takes the following form
\bea
f(E) &\,=\,& \int dx^+dx^-  e^{iE \left(\frac{x^++x^-}{2}\right)} \int d^{d-2}x \notag \\
&& \qquad \qquad \times \int dx_1^-\left[\lim_{x_1^+ \to \infty} \left(\frac{x_1^+-x_1^-}{2}\right)^{d-2} \epsilon^i \epsilon^j \langle J_i(\vect{x})T_{--}(x_1^+,x_1^-)J_j(0)\rangle \right]\,, 
\label{ed4}
\eea
where now $i,j$ will take the values $\{+,-,2\}$.
Now we use the three point correlator $\langle TJJ\rangle$ in position space to evaluate \eqref{ed4}. As discussed in Refs.~\onlinecite{Osborn:1993cr,Erdmenger:1996yc}, using the conformal symmetry and Ward identities, the form of the three point functions in $d$-dimensional CFTs can be fixed to
\bea
\langle T_{ij}(\vect{x_1})J_k(\vect{x_2}) J_l(\vect{x_3})\rangle \,=\, \frac{t_{ijmn}(\vect{X_{23}})\,g^{mp}\,g^{nq}\,I_{kp}(\vect{x_{21}})\,I_{lq}(\vect{x_{31}})}{|\vect{x_{12}}|^d\, |\vect{x_{13}}|^d\, |\vect{x_{23}}|^{d-2}}\,,
\label{ed5}
\eea
where 
\bea
\vect{x_{12}}\,=\, \vect{x_1}-\vect{x_2}\,, \quad \vect{X_{12}}\,=\,\frac{\vect{x_{13}}}{|\vect{x_{13}}|^2}-\frac{\vect{x_{23}}}{|\vect{x_{23}}|^2}\,\,\textrm{and}\,\,\hat{X}_i\,=\,\frac{X_{i}}{\sqrt{|\vect{X}|^2}}\,.
\label{ed6}
\eea
Further, we have
\bea
I_{ij}(\vect{x})&\,=\,& g_{ij}-2\frac{x_i x_j}{|\vect{x}|^2}\,, \notag \\
t_{ijmn}(\vect{X}) &\,=\,& \hat{a}\, h^1_{ij}(\vect{\hat{X}}) g_{mn}+\hat{b} \, h^1_{ij}(\vect{\hat{X}})\, h^1_{mn}(\vect{\hat{X}}) + \hat{c}\, h^2_{ijmn}(\vect{\hat{X}}) + \hat{e} \, h^3_{ijmn}(\vect{\hat{X}}) \,. \label{ed7} \\
h^1_{ij}(\vect{\hat{X}}) &\,=\,& \hat{X}_i \hat{X}_j - \frac{1}{d} g_{ij} \notag \\
h^2_{ijmn}(\vect{\hat{X}}) &\,=\,& \hat{X}_i \hat{X}_m g_{jn}+\{i\leftrightarrow j,m\leftrightarrow n\}\notag \\
&& \qquad \qquad \qquad - \frac{4}{d} \hat{X}_i \hat{X}_{j}g_{mn}-\frac{4}{d} \hat{X}_m \hat{X}_{n}g_{ij}+\frac{4}{d^2}g_{ij}g_{mn}\,, \notag \\
h^3_{ijmn}(\vect{\hat{X}}) &\,=\,& g_{im}g_{jn}+g_{in}g_{jm}-\frac{2}{d}g_{ij}g_{mn}\,.
\eea
In the above expression, $\{i\leftrightarrow j,m\leftrightarrow n\}$ represents three terms that we get by permuting the indices. Moreover in \eqref{ed7}, all the coefficients with `hat' are not independent and we have the following relations between them,
\be
d\,\hat{a}-2\hat{b}+2(d-2)\hat{c}\,=\,0\,,\qquad \hat{b}-d(d-2)\hat{e}\,=\,0\,.
\label{ed8}
\ee

Now to evaluate \eqref{ed4}, it is convenient to assume that the spacetime is even dimensional. This assumption will allow us to use the residue theorem to evaluate certain integrals when we are doing the calculation for arbitrary $d$. However, our final results are insensitive to the parity of the spacetime dimension and in the end, we can analytically continue the results to odd spacetime dimensions. Now for even $d$, we go through the following steps to compute \eqref{ed4}:
\begin{itemize}
\item First we use \eqref{ed5} to find the form of $\langle J_i(\vect{x})T_{--}(\vect{x_1}) J_j(0)\rangle$.
\item We take the limit $x_1^+\to\infty$ to get
\be
K_{i--j}\,=\,\lim_{x_1^+ \to \infty} \left(\frac{x_1^+-x_1^-}{2}\right)^{d-2} \langle J_i(\vect{x}) T_{--}(\vect{x_1}) J_j(0)\rangle
\label{ed9}
\ee
\item Next, we integrate over $x_1^-$. For that, we time order the operators using following $i\epsilon$ prescription: $x_1^0\to x_1^0-i\epsilon$ and $x^0\to x^0-2i\epsilon$.
\item We use standard results to integrate over the $(d-2)$ spatial dimensions orthogonal to $x^\pm$. 
While going through this step for different $i,j$ in \eqref{ed4}, we will find that some of the integrals are divergent. This is just an artifact of performing the integrations along the directions orthogonal to $x^{\pm}$, before integrating over $x^{\pm}$. We do so to simplify the integrations for arbitrary $d$ and to fix these spurious divergences, we use the techniques of dimensional regularization. At this step, we perform the integration over $(d-2-\kappa)$ spatial dimensions instead of $(d-2)$, and in the final result we will take the limit $\kappa\to 0$. So here, we actually calculate
\be
\int d^{d-2-\kappa}x \int dx_1^- K_{i--j}\,.
\label{ed13}
\ee
\item Now we perform the integration over $x^-$ and $x^+$:
\be
\int dx^-\,dx^+\, e^{i\frac{E}{2}x^-}e^{i\frac{E}{2}x^+} \int d^{d-2-\kappa}x \int dx_1^- K_{i--j}\,.
\label{ed14}
\ee
In the contour integrations at this step, we close the loop from above because only then the integrations will converge.
\item Finally, we take the limit $\kappa\to 0$ to get a finite result:
\be
Q_{i--j}=\lim_{\kappa \to 0} \int dx^-\,dx^+\, e^{i\frac{E}{2}x^-}e^{i\frac{E}{2}x^+} \int d^{d-2-\kappa}x \int dx_1^- K_{i--j}\,.
\label{ed14a}
\ee
\end{itemize}
We repeat the above steps for all the values of $i$ and $j$ in \eqref{ed4}. Details of these calculations can be found in attached Mathematica program \cite{energydensity} and we find that
\bea
Q_{----} &\,=\,& \frac{(d-2) \Big( (d+1) (2 d \hat{a}+(d-2) \hat{b}+4(d-2)\hat{c})+2 d (d+2) \hat{e} \Big) \pi ^{\frac{d}{2}+2}}{2^{d-1}\,(d+2)\, \Gamma\left(\frac{d+2}{2}\right)^3} \left(\frac{E}{2}\right)^{d-1} \,, \notag \\
Q_{---+} &\,=\,& -\frac{ d \Big((d-2) (d+1) \hat{b}+d (d \hat{a}+2(d-2) \hat{c})\Big) \pi^{\frac{d}{2}+2}}{2^d\,\Gamma \left(\frac{d}{2}+2\right) \Gamma \left(\frac{d}{2}+1\right)^2} \left(\frac{E}{2}\right)^{d-1} \,, \notag \\
Q_{---2} &\,=\,& 0\,, \notag \\
Q_{+---} &\,=\,& -\frac{ \left((d \hat{a}+2(d-2) \hat{c}) \Gamma \left(\frac{d}{2}-1\right) \Gamma\left(\frac{d}{2}+1\right)+(d+1) \hat{b} \Gamma \left(\frac{d}{2}\right)^2\right)\pi^{\frac{d}{2}+2} }{2^{d-2}\, \Gamma \left(\frac{d}{2}-1\right) \Gamma \left(\frac{d}{2}+2\right) \Gamma \left(\frac{d}{2}\right)^2 \Gamma \left(\frac{d}{2}+1\right)}  \left(\frac{E}{2}\right)^{d-1} \,, \notag \\
Q_{+--+} &\,=\,& \frac{ d(d-1)\,\hat{b}\, \pi ^{\frac{d}{2}+2}}{2^{d-1}\, \Gamma \left(\frac{d}{2}+1\right)^3}  \left(\frac{E}{2}\right)^{d-1} \,, \notag \\
Q_{+--2} &\,=\,& 0 \,, \notag \\
Q_{2---} &\, =\,& 0 \,, \notag \\
Q_{2--+} &\, =\,&0 \,, \notag \\
Q_{2--2} &\, =\,& -\frac{ (d \hat{a}-4 \hat{c}) \pi ^{\frac{d}{2}+2}}{2^{d-3}\,\Gamma \left(\frac{d}{2}\right) \Gamma \left(\frac{d}{2}+1\right)^2} \left(\frac{E}{2}\right)^{d-1} \,. \notag
\eea
Using these values for $Q_{ijkl}$ and relations \eqref{ed8} in \eqref{ed4}, we find that the energy flux for arbitrary $d$ becomes
\be
\langle \mathcal{E}(\vect{n})\rangle \,=\, \frac{E}{\Omega_d} \left(1 - \frac{(d-1)\left(d(d-2)\hat{e}-\hat{c}\right)}{(d-2)\left(\hat{e}+\hat{c}\right)}\left(\cos^2\theta -\frac{1}{d-1} \right) \right)\,.
\label{ed15}
\ee
Note that the two point function in the denominator of \eqref{ed1} does not have any angular dependence, and it fixes the normalization of higher point functions. Now, we can easily read off the value of $\mathcal{A}$ from \eqref{ed15} and also find it to be consistent with results for $d=4$ in Ref.~\onlinecite{Hofman:2008ar}.

In Ref.~\onlinecite{Osborn:1993cr}, Osborn and Petkou have further studied the position three-point functions for the specific conformal field theory \eqref{lcft}. By calculating the collinear three point functions $\langle TJJ\rangle$ for free scalar and free fermions, they have found the ratio of the coefficients $\hat{c}$ and $\hat{e}$ to be
\be
\left( \frac{\hat{e}}{\hat{c}} \right)_{s}\,=\,\frac{1}{d-2} \quad \textrm{and}\quad \left( \frac{\hat{e}}{\hat{c}} \right)_{f}\,=\,0\,.
\label{ed16}
\ee
These can be further used to find the value of $\mathcal{A}$ in scalar and fermionic conformal field theories to be
\be
\mathcal{A}_s \,=\, d-1 \quad \textrm{and} \quad \mathcal{A}_f\,=\, -\frac{d-1}{d-2}\,. 
\label{ed17}
\ee
In the next section, we show how $\mathcal{A}$ is related to the coupling constant $\gamma$ in action \eqref{Smyers} for $d=3$. Then, these results are compared with the CFT results \eqref{ed17} to fix $\gamma$ for free field theories.

\subsection{$\mathcal{A}$ from holography and matching the results}

The holographic computation of $\mathcal{A}$ for $d=4$ was first done in Ref.~\onlinecite{Hofman:2008ar} and then was extended to $d=3$ in Ref.~\onlinecite{Myers:2010pk}. These calculations can be easily generalized to arbitrary dimensions and we find that
\be
\mathcal{A}\,=\,-4\,d(d-1)\gamma\,.
\label{ed18}
\ee
A quick overview of the holographic computation is as follows. According to the AdS/CFT dictionary, the computation of expectation value of energy flux in the boundary theory, for a state created by a conserved vector current, boils down to calculating the three point function between two photons and a graviton. To compute such a three point function in the bulk gravity \eqref{Smyers}, we need to introduce appropriate metric fluctuations and two gauge field perturbations in the $(d+1)$-dimensional AdS background. These fluctuations couple to the stress-energy tensor and vector current insertions $T_{ij}$ and $J_i$ on the boundary and one needs to evaluate their on-shell contribution for the action \eqref{Smyers}, as was done in Section~\ref{sec:holo3}. The bulk action has two terms. We find that the first term only contributes to the angle independent component of \eqref{ed2} and the second term introduces the anisotropy in the flux distribution. Hence, merely by comparing the contributions from both of the terms, we can easily extract the coefficient $\mathcal{A}$. For more details of this calculation, interested readers can refer to Appendix D of Refs.~\onlinecite{Hofman:2008ar} and \onlinecite{Myers:2010jv}.

Now we match the field theory and holographic calculations from Eqs.~\eqref{ed17} and \eqref{ed18} for $d=3$ to find following values of $\gamma$ for free scalars and fermions
\be
\gamma_s\,=\,-\frac{1}{12}\quad \textrm{and} \quad \gamma_{f}\,=\,\frac{1}{12}\,,
\ee
which indeed are consistent with the momentum space calculations in \eqref{gammas} and \eqref{gammaf}
in the limit $N_F = \infty$.

\section{Conclusions}
\label{sec:conc}

The primary results of this paper are the values of $\gamma$ in Eq.~(\ref{gammares}) for the conserved currents of the \bl{2+1 dimensional} CFTs defined
in (\ref{lcft}). Here $\gamma$ is defined as a parameter controlling the structure of the zero temperature 
three-point correlator $\langle T J J \rangle$ between
the stress-energy tensor and the conserved current. 
\bl{Osborn and Petkou \cite{Osborn:1993cr}
specified the general form of the $\langle T J J \rangle$ correlator, and $\gamma$ was exactly connected to their
parameterization in Section~\ref{sec:ajay}.}
However, $\gamma$ also appears 
in the holographic representation of the CFT on AdS$_4$, and is the coupling constant determining a four-derivative term in a gradient expansion
of the effective action: see Eq.~(\ref{Smyers}). The latter connection endows $\gamma$ with much greater physical importance:
it determines the structure of a variety of dynamical properties of charge transport at non-zero temperatures, both equilibrium and 
non-equilibrium. \bl{The holographic formulation also leads to the bound $|\gamma| \leq 1/12$ \cite{Myers:2010pk}}.

The action (\ref{Smyers}) was derived in Ref.~\onlinecite{Myers:2010pk} as the most general 4-derivative holographic theory in an effective field theory framework, which is expressed in terms of the gauge flux $F_{\mu\nu}$ and the metric tensor. 
Further, it is explicitly shown in appendix \ref{app:action} that the holographic computation of three-point function $\langle TJJ\rangle$ is independent of the choice of four-derivative terms in the action if we reparameterize the couplings properly. 
In a string theory context, higher-derivative interaction terms in this action are suppressed by the ratio of string length scale over the curvature scale of background geometry. For our perspective, we are viewing the effective field theory as one in which loop corrections
are already included in the values of the couplings, and so is to be evaluated only at tree level. Corrections to our analysis arise from 6 (and higher) derivative terms, and we have not established that such terms are quantitatively small. However, it is encouraging to note that
the 4-derivative corrections to the 2-derivative conductivity were smaller than 33\% at all frequencies, 
and this was in turn related to the bound on $\gamma$ \cite{Myers:2010pk}. Also reassuring is the fact that the holographically obtained bound $|\gamma| \leq 1/12$ coincides with 
exact bound obtained from CFT methods in Section~\ref{sec:ajay}. 

It will be interesting to push this phenomenological approach by augmenting the action \eqref{Smyers} by other fields, which are holographic duals of other primary operators of the CFT \cite{quasinormal,JMthanks}. 
The most important of these is the ``mass'' term $|z_a|^2$ in (\ref{lcft}), which tunes
the CFT away from the critical point at $T=0$. Here we are assuming we are at the CFT critical point at $T=0$, and so such a relevant
perturbation is not present in the underlying theory at $T=0$; \bl{the structure of the interactions in the CFT ensures that there is no change in $\langle |z_a|^2 \rangle$ at $T>0$ \cite{CSY}.}
In the holographic theory, $|z_a|^2$ is represented by a scalar dilaton field, $\bl{\Phi}$.
This can influence charge transport by an additional term $\sim \bl{\Phi} F_{\mu\nu} F^{\mu\nu}$ in (\ref{Smyers}). Such a $\bl{\Phi}$ does
not have an expectation value in the AdS$_4$ theory 
at $T=0$, and will not acquire one at $T>0$ in the absence of external sources. In the linear response
computation of the conductivity from such an augmented action, the $\sim \bl{\Phi} F_{\mu\nu} F^{\mu\nu}$ term only influences
the conductivity at the one-loop level in the bulk theory, so need not be included in our tree-level treatment of the effective theory
(\ref{Smyers}). Thus $\gamma$ remains as the crucial coupling determining the structure of the charge transport properties of the CFT,
as was noted recently \cite{quasinormal}.

In Refs.~\onlinecite{Myers:2010pk,quasinormal}, it was shown that $\gamma$ determined the structure of the universal frequency dependence of the
conductivity $\sigma (\omega)$ at non-zero temperatures. 
For $0 < \gamma \leq 1/12$, it was found
that there was a Drude-like peak at $\omega = 0$, followed by an eventual saturation at a constant at large $\omega$.
Such a structure appears physically reasonable from our present computation of $\gamma = 1/12$ for the free-fermion theory with $N_s=0$:
the free fermion theory has a delta function at zero frequency \cite{Sachdev:2010uz}, and it is expected that this will be broadened
to a Drude peak upon including interactions. 

In the complementary range $-1/12 \leq \gamma < 0$, it was found \cite{Myers:2010pk,quasinormal} that $\sigma (\omega)$ had
a `dip' at $\omega = 0$, rather than a peak. The value $\gamma =-1/12$ is obtained for the free scalar theory with $N_f = 0$.
We can understand this dip if we interpret the scalar field in (\ref{lcft}) as representing a {\em vortex\/} degree of freedom near {\em e.g.\/}
a superfluid-insulator quantum phase transition \cite{Sachdev:2010uz}. Particle-vortex duality maps the conductivity to its inverse,
and the inverse conductivity then has a Drude-like peak at $\omega = 0$.
Further evidence for this interpretation comes from our computation of $\gamma_t  = 1/12$ obtained with $N_f=0$ for the topological 
current of (\ref{lcft}). Under particle-vortex duality, the charged particle current
in the dual theory maps to the topological current of (\ref{lcft}), and so this also implies a peak in 
$\sigma (\omega)$ for the charged particle current.

Further applications include computation of other dynamical consequences of the value of $\gamma$. In a recent work \cite{quasinormal},
it was shown that $\gamma$ crucially determined the structure of the poles and zeros of the complex conductivity in the lower-half
of the complex frequency plane. These poles and zeros are associated with quasinormal modes of the holographic theory,
and they are expected to be central to an understanding of the thermal dynamics of the CFT.
Combined with more precise computations of the value of $\gamma$ by the methods of the present paper, these connections
open up the possibility of precise predictions for the dynamics of the strongly-interacting condensed matter systems.

\acknowledgments

We thank I.~Feige, D.~Hofman, J.~Hung, J.~Maldacena, P.~McFadden, R.~Myers, A.~Petkou, M.~Smolkin, and W. Witczak-Krempa for useful
discussions. We are particularly grateful to D.~Hofman for providing details of the computation
in Ref.~\onlinecite{Hofman:2008ar}, and helping track down a numerical error
in the computation of Section~\ref{sec:ajay}. S.R.
is partially supported by a Ramanujan fellowship of the Department of
Science and Technology (India). S.R. is grateful to the Harvard University physics department for its hospitality while this work was being completed. D.C., S.S., and P.S.
are partially supported by the U.S. National Science Foundation under grant DMR-1103860, and by 
the U.S. Army Research Office Award W911NF-12-1-0227. P.S. 
also acknowledges support from the Deutsche Forschungsgemeinschaft under grant Str 1176/1-1. Research at Perimeter Institute is supported by the Government of Canada through Industry Canada and by the Province of Ontario through the Ministry of Research \& Innovation.

\appendix

\section{Holographic action and higher derivative interactions}
\label{app:action}

In this appendix, we discuss the effective holographic action for the gauge field $A_\mu$ in bulk gravity, and four-derivative corrections. We first recall the discussion in Ref.~\onlinecite{Myers:2010pk} where the most general four-derivative effective action 
of holographic theories was presented.
Here we add further details to this argument by analyzing the full parameter space of four-derivative interactions of holographic theories, and show that it reduces to \eqref{Smyers} for the case of linear charge transport in CFTs at zero charge density in
2+1 dimensions.

As discussed in Ref.~\onlinecite{Myers:2010pk}, generally the interaction terms in the bulk gravity action are organized by the number of derivatives. For a gauge field in four-dimensional bulk gravity, one can construct 15 covariant and parity conserving four-derivative terms from gauge field, metric and their derivatives. One can further use integration by parts and identities like $\nabla_{[\mu}F_{\nu\sigma]}=0=R_{[\mu\nu\sigma]\rho}$ and reduce the action to have 8 independent four-derivative terms:
\bea
I_{vec}&=& \frac{1}{g_4^2}\int d^4x \sqrt{-g}\bigg[ -\frac{1}{4}F_{\mu\nu}F^{\mu\nu} + L^2\,\big[\alpha_1 R^2+\alpha_2 R_{\mu\nu}R^{\mu\nu}+ \alpha_3 \left(F^2\right)^2+ \alpha_4 F^4  \nonumber\\
&& \qquad\quad + \alpha_5 \nabla^\mu F_{\mu\nu} \nabla^\sigma F_\sigma{}^\nu+ \alpha_6 R_{\mu\nu\sigma\rho}F^{\mu\nu}F^{\sigma\rho} + \alpha_7 R^{\mu\nu}F_{\mu\sigma}F_\nu{}^\sigma + \alpha_8 R F^2 \big] \bigg]\,,
\label{new1x1}
\eea
where $F^2=F^{\mu\nu}F_{\mu\nu}$, $F^4=F^\mu_\nu F^\nu_\sigma F^\sigma_\rho F^\rho_\mu$ and $\alpha_i$ are some unspecified, dimensionless constants. In the context of string theory, we can expect these interactions to appear in the low-energy effective action as string loop or $\alpha'$ corrections to the leading two-derivative action \cite{Hanaki:2006pj,Cremonini:2008tw}. Hence, these new interactions will be part of a perturbative expansion where the contribution of the higher derivative terms will be suppressed by powers of string scale over the curvature scale of background. In this framework, we can also use field redefinitions to set all the coupling constants, excluding $\alpha_3$, $\alpha_4$ and $\alpha_6$, to zero \cite{Myers:2009ij}. Now in the remaining action, $\alpha_3$ and $\alpha_4$ interaction contain four powers of field strength. Hence, these terms do not contribute to the three-point function $\langle TJJ \rangle$ and to the linear charge transport properties of the CFT. So in our effective field theory framework, the only relevant terms needed for a phenomenological comparison of charge transport properties with CFTs are the $\alpha_{6,7,8}$ interactions. In the action \eqref{Smyers}, the contribution of these 
four-derivative interactions are formulated in terms of Weyl tensor.

Now we focus on the relevant terms of the most general four-derivative action \eqref{newx1}, which contribute 
to the $\langle TJJ \rangle$ correlator:
\begin{equation}
I'_{vec}=\frac{1}{\tilde{g}^2_4} \int d^4x \sqrt{-g}\left[-\frac{1}{4}F_{\mu\nu}F^{\mu\nu} + L^2 \left[\alpha_6 R_{\mu\nu\sigma\rho}F^{\mu\nu}F^{\sigma\rho} + \alpha_7 R_{\mu\nu}F^{\mu\sigma}F^\nu{}_{\rho} + \alpha_8 R F^2 \right] \right]\,.
\label{newx1}
\end{equation}
Note that for particular relative values of $\alpha_6$, $\alpha_7$ and $\alpha_8$, this action takes the form of \eqref{Smyers} using \eqref{newx2}. However as we show now, even for arbitrary couplings, all the charge transport properties of this new model are identical to \eqref{Smyers} for a CFT at zero density. 
For the action \eqref{newx1}, we can find the equations of motion and confirm that AdS vacuum and neutral black hole solutions remain unmodified. Particularly, the black hole solution satisfies the vacuum Einstein equations $R_{\mu\nu}=-3/L^2g_{\mu\nu}$. Further, the Riemann curvature tensor $R_{\mu\nu\sigma\rho}$ is related to the Weyl tensor $C_{\mu\nu\sigma\rho}$ by the following relation
\be
R_{\mu\nu\sigma\rho}=C_{\mu\nu\sigma\rho}+g_{\mu[\sigma}R_{\rho]\nu}-g_{\nu[\sigma}R_{\rho]\mu}-\frac{1}{3}R\,g_{\mu[\sigma}g_{\rho]\nu}\,.
\label{newx2}
\ee
By substituting these relations into the action \eqref{newx1}, we find that the action becomes
\be
I'_{vec}=\frac{1+8\alpha_6+12\alpha_7 + 48\alpha_8}{\tilde{g}^2_4} \int d^4x \sqrt{-g}\left(-\frac{1}{4}F_{\mu\nu}F^{\mu\nu} + \frac{\alpha_6}{1+8\alpha_6+12\alpha_7 + 48\alpha_8} L^2 C_{\mu\nu\sigma\rho}F^{\mu\nu}F^{\sigma\rho} \right)\,.
\label{newx3}
\ee
Hence, expression \eqref{newx1} of the action is identical to \eqref{Smyers} if we define 
\be
g_4^2=\frac{\tilde{g}^2_4}{1+8\alpha_6+12\alpha_7 + 48\alpha_8}\quad{\rm and}\quad \gamma=\frac{\alpha_6}{1+8\alpha_6+12\alpha_7 + 48\alpha_8}\,.
\label{newx4}
\ee
This implies that all the charge transport properties of neutral CFTs in the generalized theory \eqref{newx1} are the same as that of \eqref{Smyers} with a proper reparametrization. To further support this argument, we can find the bounds on couplings $\alpha_6$, $\alpha_7$ and $\alpha_8$ by directly applying the procedure in Section 5 of \cite{Myers:2010pk} on action \eqref{newx1}. The values of these couplings are constrained by demanding that the CFT dual to bulk gravity \eqref{newx1} respects causality \cite{Buchel:2009tt,Brigante:2008gz,Ritz:2008kh}, and there are no unstable modes of vector field \cite{Myers:2007we,Brigante:2007nu}. For our four derivative action at tree level, we find that $|{\alpha_6}/{(1+8\alpha_6+12\alpha_7 + 48\alpha_8)}| \leq 1/12$,  which is consistent with \eqref{newx4} and the bound $|\gamma|\leq 1/12$. Although here we can not fix the numerical values of couplings $\alpha_6$, $\alpha_7$ and $\alpha_8$, our results in this paper for 
a CFT at zero density are independent of choice of four-derivative terms in the action.

\section{Review of the CFT two-point correlators $\langle J J \rangle$ and $\langle T T \rangle$}
\label{app:two_point}

In this appendix, we derive the current and stress-tensor two-point functions given in Eq.~(\ref{ej}, \ref{et}) and compute 
$C_J$ and $C_T$ for the free theory. In momentum space, the two point function for currents and for the stress-tensor is just given by two bubble diagrams: one with two scalar boson propagators and the other with  two fermion propagators, respectively.
The scalar boson contribution to the current-current correlator reads:
\be
\ep_{1,i_1} \ep_{2,i_2} \langle J_s^{i_1}(-{\bf k}) J_s^{i_2}({\bf k}) \rangle = 4 N_s \ep_{1}^{i_1} \ep_{2}^{i_2} \int {P_{i_1} P_{i_2} \over P^2 (\vect{P} + \vect{k})^2} {d^3 P \over 8 \pi^3}.
\ee
Using the identity
\be
\int {P_{i_1} P_{i_2} \over P^2 (\vect{P} + \vect{k})^2} {d^3 P \over 8 \pi^3} = \bigg(3 {k_{i_1} k_{i_2} \over \norm{k}^2} - \eta_{i_1 i_2}\bigg) {\norm{k} \over 64},
\label{2ptjj}
\ee
one obtains
\be
\ep_{1,i_1} \ep_{2,i_2} \langle J_s^{i_1} J_s^{i_2} \rangle = -N_s {\norm{k} \over 16} \vect{\ep_1} \cdot \vect{\ep_2}, 
\ee
in agreement with the uncontracted expression in \eqref{ej}. This yields $C_{J,s}=N_s/16$ for the complex scalars. The calculation for free Dirac fermions is similar and one obtains $C_{J,f}=N_f/16$. 

For the two point function of the stress-tensor, the scalar boson bubble can be integrated using the identity
\be
\begin{split}
\int \frac{d^3 P}{8 \pi^3} \frac{P_{i_1} P_{i_2} P_{i_3} P_{i_4} }{P^2 ({\vect{P}} + {\vect{k}})^2} = 
&( \eta_{i_1i_2} \eta_{i_3i_4} + \mbox{2 terms} ) \frac{\norm{k}^3}{1024}
- \left( \frac{k_{i_1} k_{i_2}}{\norm{k}^2}  \eta_{i_3 i_4} + \mbox{5 terms} \right) \frac{5 \norm{k}^3}{1024}
\\ &+ \frac{35 k_{i_1} k_{i_2} k_{i_3} k_{i_4}}{1024 \norm{k}},
\label{2pttt}
\end{split}
\ee
resulting in the expression 
\be
\ep_{1,i_1} \ep_{1,i_2} \ep_{2,i_3} \ep_{2,i_4} \langle T_s^{i_1 i_2}(\vect{k}) T_s^{i_3 i_4}(-\vect{k}) \rangle = 
{N_s \over 256} 2 (\vect{\ep_1} \cdot \vect{\ep_2})^2 \norm{k}^3\;.
\ee
Note that this agrees with \eqref{et} with $C_{T,s}=N_s/256$. An identical computation for the two point function of the stress-tensor for free Dirac fermions gives $C_{T,f}=N_f/256$. 

These quantities enable us to fix the values of certain coupling constants in the gravity theory in Section \ref{sec:matching}.

\section{AdS/CFT Correlators and Two Point Functions}
\label{sec:2point}

This appendix provides background on the methods of gauge-gravity duality for readers who are condensed matter physicists.

The AdS/CFT conjecture states that theories of quantum gravity on $d+1$ dimensional anti-de Sitter space (denoted \ads[d+1]), are dual to $d$-dimensional conformal field theories that live on the ``boundary'' of AdS. The theory in \ads[d+1] is called the ``bulk theory'' and the theory on the boundary is called the ``boundary theory.'' In the version of the correspondence that we will be using here, the bulk theory will live on the ``Poincare patch'' of \ads[d+1] (the metric for this patch was already described above), while the boundary theory will live on $R^{d-1,1}$. 

More precisely, the conjecture states that each ``field'' in the bulk corresponds to an operator on the boundary; second, if we do the path integral in the bulk theory with asymptotic boundary conditions fixed for these fields, then this equals the generating functional of the boundary theory with sources turned on for the corresponding operators.

Such an approach makes sense as long as we can distinguish individual fields in the bulk, and there is a corresponding decomposition of the spectrum of operators in the boundary theory in terms of single and double trace operators. This decomposition is possible for theories with a large-N expansion, and it is in this regime that the AdS/CFT conjecture has been widely tested.

We now describe this conjecture quantitatively and explain how it may be used to calculate and compare correlation functions. 

\subsection{Prescription}
In this section, we describe the prescription for computing correlation functions in AdS/CFT. This follows Ref.~\onlinecite{Witten:1998qj} with some refinements that were made in Ref.~\onlinecite{Freedman:1998tz}. (See Ref.~\onlinecite{Aharony:1999ti} for a review.)

A scalar field of mass-squared $m^2$ in the bulk is dual to an operator of dimension $\Delta = {d \over 2} + \sqrt{\left({d \over 2}\right)^2 + m^2}$ on the boundary. If we solve the equations of motion for a free field of this mass we find, that near the boundary, we can have $\phi \sim z^{d - \Delta}$ and $\phi \sim z^{\Delta}$. The solution that grows at the boundary is called the ``non-normalizable'' solution, while the other one is called the ``normalizable'' solution. If we work in Euclidean AdS, then fixing the coefficient of the non-normalizable mode, and demanding regularity in the interior automatically fixes the normalizable mode also. In Lorentzian AdS, the normalizable mode can be set independently for time-like momenta, but below we will consider those solutions that come from a continuation of the Euclidean solutions.

The original prescription for correlation functions \cite{Witten:1998qj} was given for massless fields. For massive fields, we need to be careful about regularization because the non-normalizable mode diverges as we approach the boundary. 
So, we will cut the AdS space off at $z = \epsilon$, and consider doing the bulk path integral with the following regularized boundary condition for the scalar field as we approach the boundary:
\be
\label{boundcndns}
\phi(\vect{x}, z) \underset{z \rightarrow \epsilon}{\longrightarrow} \epsilon^{d - \Delta} \phi_0(\vect{x}).
\ee
The idea is to work with this boundary condition and extract the finite part in $\epsilon$ at the end of the calculation.
Then, the AdS/CFT prescription is that:
\be
\label{adscftpres}
\left. \int e^{-S} {\cal D} \phi\right|_{\text{bound}} = \langle e^{\int \phi_0(\vect{x}) O(\vect{x}) d^d \vect{x}} \rangle_{\text{CFT}}.
\ee
Here the left hand side is short hand for the path integral in the bulk done with the boundary conditions \eqref{boundcndns} while the right hand side is an expectation value in the conformal field theory. Although, to lighten the notation, we have chosen a particular coordinate system to represent the boundary conditions \eqref{boundcndns}, the prescription is independent of this choice. 

The original conjecture \eqref{adscftpres} was made in specific contexts: for example, one of the best studied examples of the AdS/CFT duality is when the bulk theory is type IIB string theory on $\text{AdS}_5 \times S^5$ and the boundary theory is ${\cal N}=4$ super-Yang-Mills theory. Several other examples are known.

However, where correlation functions are concerned, the prescription \eqref{adscftpres} may be examined just as well within effective field theory. This means that we take some effective field theory in the bulk and compute the left hand side of \eqref{adscftpres} at tree-level in the bulk. This computation can be used to {\em define} a generating functional in a CFT to leading order in ${1 \over N}$.\footnote{As we mentioned above, the prescription \eqref{adscftpres} makes sense when we have a perturbative parameter that allows us to differentiate between single and double trace operators, and we are using $N$ as a short-hand for this parameter here.}
 This is because one can show that the quantity obtained this way satisfies all the constraints of conformal invariance and the operator product expansion (OPE) to leading order in ${1 \over N}$ in the boundary theory.

Now let us turn to the stress-tensor and conserved currents.
The graviton in the bulk is dual to the stress tensor on the boundary, and a gauge field is dual to conserved currents. Now, consider doing the bulk path integral with the following boundary conditions for the metric and the gauge fields:
\be
\label{boundcndnsgraviton}
\begin{split}
&g_{z z}(\vect{x}, z) \underset{z \rightarrow 0}{\longrightarrow} {1 \over z^2}; g_{z i}(\vect{x},z) \underset{z \rightarrow 0}{\longrightarrow} 0; g_{i j}(\vect{x}, z) \underset{z \rightarrow 0}{\longrightarrow}  {1 \over z^2} \left(\eta_{i j} + \chi_{i j}(\vect{x}) \right), \\
&A_{z}(\vect{x},z)  \underset{z \rightarrow 0}{\longrightarrow}  0; A_{i}(\vect{x}, z) \underset{z \rightarrow 0}{\longrightarrow} \mathbb{V}_{i}(\vect{x}).
\end{split}
\ee
Then the bulk path integral with these boundary conditions is conjectured to be the same as the following generating functional of the conformal field theory:
\[
\langle e^{\int \left[\chi_{i j}(\vect{x}) T^{i j}(\vect{x}) + \mathbb{V}_{i}(\vect{x}) j^{i}(\vect{x}) \right] d^d \vect{x} } \rangle .
\]

\subsection{Scalar Two Point Function}
The simplest setting in which we can test these ideas is to evaluate two-point functions. Consider a free massive scalar with action:
\be
S_{\text{bulk}} = -{1 \over 2} \int \sqrt{-g} \left[(\partial_{\mu} \phi)^2 + m^2 \phi^2\right] .
\ee
At leading order we can evaluate the left hand side of \eqref{adscftpres} in the saddle point approximation. Let us also take 
\be
\phi_0(\vect{x}) = \lambda_1 e^{i \vect{k_1} \cdot \vect{x}} + \lambda_2 e^{i \vect{k_2} \cdot \vect{x}}.
\ee
We need to find a solution of the equations of motion:
\be
(\Box - m^2) \phi = 0,
\ee
that respects \eqref{boundcndns}. 

In fact, it is rather subtle to write down such a solution. The authors of Ref.~\onlinecite{Freedman:1998tz} showed that the correct method is to write down the following solution:
\be
\label{solphitwomom}
\phi(\vect{x}, z) =  \epsilon^{d - \Delta} \left[\la_1  {(\norm{k_1} z)^{d \over 2} K_{\Delta - {d \over 2}}(\norm{k_1} z)  \over (\norm{k_1} \epsilon)^{d \over 2} K_{\Delta - {d \over 2}}(\norm{k_1} \epsilon)} e^{i \vect{k_1} \cdot \vect{x}} + \la_2 {(\norm{k_2} z)^{d \over 2} K_{\Delta - {d \over 2}}(\norm{k_2} z) \over (\norm{k_2} \epsilon)^{d \over 2} K_{\Delta - {d \over 2}} (\norm{k_2} \epsilon)} e^{i \vect{k_2} \cdot \vect{x}}  \right],
\ee
where $K$ is the modified Bessel function. Here we have defined $\norm{k_m}$ to be taken in the Lorentzian metric, with a mostly positive signature i.e the boundary metric is defined to be $\text{diag}(-1,1,1 \ldots 1)$.  For timelike $\vect{k}$, we should take its norm to have a negative imaginary part; this continues the modified Bessel function $K$ to a Hankel function $H^{(1)}$. 

We can superpose solutions of different momenta, so that the sum has delta function support at a given point; such a solution is called a ``bulk to boundary'' propagator. If we Fourier transform the bulk to boundary propagator, we will get a solution of the sort above. 

It is very tempting to expand \eqref{solphitwomom} in powers of $\epsilon$ so that we have:
\be
\label{naivebulkbound}
\begin{split}
\phi(\vect{x}, z) &=  \frac{2^{\frac{1}{2} (d-2 \Delta ) + 1}}{\Gamma
   \left(-\frac{d}{2}+\Delta\right)}  \left[\la_1  \norm{k_1}^{\Delta -\frac{d}{2}} z^{d \over 2} K_{\Delta - {d \over 2}}(\norm{k_1} z) e^{i \vect{k_1} \cdot \vect{x}}  + \la_2 \norm{k_2}^{\Delta -\frac{d}{2}} z^{d \over 2} K_{\Delta - {d \over 2}}(\norm{k_2} z) \right] \\ &+ \Or[\epsilon^{2 \Delta - d}] + \Or[\epsilon].
\end{split}
\ee
However, as was shown in Ref.~\onlinecite{Freedman:1998tz}, we {\em cannot} discard the subleading terms in $\epsilon$ at this stage because there is a second divergence when we evaluate the on-shell action and these subleading terms then contribute at $\Or[\epsilon^0]$ in the final answer.

Now, let us compute the two point function using the prescription above. The on-shell action is divergent if we take $\epsilon \rightarrow 0$, so we should do the calculation with $\epsilon$ kept finite and extract the $\epsilon^0$ term at the end. 

On the solution \eqref{solphitwomom}, the on-shell action is simply:
\be
\label{onshell}
S_{\text{on-shell}} = {-1 \over 2} \left. \int \sqrt{-g} z^2 \phi {\partial \phi \over \partial z} d^d x \right|_{z = \epsilon}.
\ee
A short calculation shows that the $\epsilon^0$ term on the right hand side of \eqref{onshell} that is bilinear in $\lambda_1$ and $\lambda_2$ is:
\be
\label{doublederivativeaction}
{\partial^2 S_{\text{on-shell}} \over \partial \la_1 \partial \la_2} = -(2 \Delta - d){\Gamma({d \over 2} + 1 - \Delta) \over \Gamma(\Delta - {d \over 2} + 1)}
   \left({\norm{k_1} \over 2}\right)^{2 \Delta - d} \delta(\vect{k_1} + \vect{k_2}) + \ldots,
\ee
where the $\ldots$ are higher and lower order terms in $\epsilon$. The terms that are divergent as $\epsilon \rightarrow 0$, are analytic in the momentum, and so they can be removed by local counterterms.

From the prescription, \eqref{adscftpres}, we can see that this is also the two point function of the operator $O$ in the conformal field theory:
\be
\langle O(\vect{k_1}) O(\vect{k_2}) \rangle = C_{\Delta} \norm{k_1}^{2 \Delta - d} \delta(\vect{k_1} + \vect{k_2}),
\ee
where $C_{\Delta}$ is the numerical constant in \eqref{doublederivativeaction}. 
In fact, this is precisely what one expects from conformal invariance,
for a primary operator of dimension $\Delta$.

\subsection{An Alternate Prescription}
\label{app:alternate}

For the stress-tensor, and even for scalar fields,  at leading order in ${1 \over N}$ (i.e at tree level in the bulk), it is often convenient to replace the prescription \eqref{adscftpres} with an equivalent prescription\cite{Banks:1998dd}. This prescription simply states that if we write the metric as:
\be
g_{\mu \nu} = g_{\mu \nu}^{\text{AdS}} + h_{\mu \nu},
\ee
where $g_{\mu \nu}^{\text{AdS}}$ is the metric \eqref{metricpoincare}, 
and consider field configurations that satisfy the asymptotic conditions \eqref{boundcndns} then:
\be
\label{alternateadscft}
\langle T_{i_1 j_1}(\vect{x_1}) \ldots T_{i_n j_n}(\vect{x_n}) \rangle_{\text{boundary}} = Z^n \lim_{z_i \rightarrow 0} z_1^{2-d} \ldots z_n^{2-d} \langle h_{i_1 j_1}(\vect{x_1}, z_1) \ldots h_{i_n j_n}(\vect{x_n}, z_n) \rangle_{\text{bulk}}.
\ee
This is the statement that:
{\em boundary correlators are just boundary values of bulk Green's functions.}
Here $Z$ is a wave-function renormalization factor. At tree-level in the bulk, this factor is just a constant as we will see below, and so we have written $Z^n$ rather than writing separate factors for each insertion. $Z$ just fixes the overall normalization of operators and so, at tree-level, it is not physically relevant but we will retain it for later convenience.
For scalar operators, the analogous prescription is:
\be
\langle O(\vect{x_1}) \ldots O(\vect{x_n}) \rangle_{\text{boundary}} = Z^n \lim_{z \rightarrow 0} z_1^{-\Delta} \ldots z_n^{-\Delta} \langle \phi(\vect{x_1}, z_1) \ldots \phi(\vect{x_n}, z_n) \rangle_{\text{bulk}}.
\ee

This is the prescription that we will use to evaluate two point functions.

\subsection{Scalar Two Point Function Rederived}
To get a feel for this prescription, let us re-derive the result above for the two point function of scalar operators. The scalar two-point Green's function in the bulk is given by Ref.~\onlinecite{Liu:1998ty}:
\be
\begin{split}
G(\vect{x}, z, \vect{x'}, z') &= \int  {d^d \vect{k} \over (2 \pi)^d} G_{\vect{k}}(z,z') e^{-i \vect{k} \cdot (\vect{x} - \vect{x'})} \\
&= -\int {d^d \vect{k} \over (2 \pi)^d} {d p^2 \over 2}  {e^{i \vect{k} \cdot (\vect{x} - \vect{x'})} 
z^{d \over 2} J_{\Delta - {d \over 2}}(p z) J_{\Delta - {d \over 2}} (p z') (z')^{d \over 2} \over 
\left(\vect{k}^2 + p^2 - i \epsilon \right)} .
\end{split}
\ee
We can check that this Green's function obeys:\footnote{Note that Ref.~\onlinecite{Liu:1998ty} defines the Green's function with an additional minus sign on the right hand side.}
\be
\label{greenequation}
(\Box - m^2)G(\vect{x}, z, \vect{x'}, z') = {1 \over \sqrt{-g}} \delta(\vect{x} - \vect{x'}) \delta(z - z').
\ee
In Fourier space, the relation \eqref{greenequation} is simply:
\be
z^{d + 1} {\partial \over \partial z} z^{1 - d} {\partial G_{k}(z, z') \over \partial z} - m^2 G_k(z,z') - z^2 k^2 G_k(z,z') = \delta(z - z') z^{d + 1}.
\ee
We can verify that this is satisfied by virtue of the identity:
\be
\int p J_{\nu} (p z) J_{\nu} (p z') d p = z^{-1} \delta(z - z').
\ee

After doing the $p$ integral and transforming to momentum space, we find that the two point Green function can be written:
\be
G(\vect{k}, z_1, z_2) = -(z_1 z_2)^{d \over 2} 
 I_{\Delta - {d \over 2}}(\norm{k} z^{<}) K_{\Delta - {d \over 2}} (\norm{k} z^{>}),
\ee
where $z^{<} = \text{min}(z_1, z_2)$ and $z^{>} = \text{max}(z_1, z_2)$.\footnote{We have written this as a function of one momentum, rather than two, because the two momenta are forced to be equal by momentum conservation.}

 With this choice, when we now take the limit where one point goes to the boundary, and also take $Z = -(2 \Delta - d)$, we find:
\be
Z \lim_{z_1 \rightarrow 0} z_1^{-\Delta} G(\vect{k}, z_1, z_2) = \frac{2^{\frac{1}{2} (d-2 \Delta ) + 1}}{\Gamma
   \left(-\frac{d}{2}+\Delta\right)}  \norm{k}^{\Delta -\frac{d}{2}} z_2^{d \over 2} K_{\Delta - {d \over 2}} (\norm{k} z_2).
\ee
Note that this matches the ``naive'' bulk to boundary propagator of \eqref{naivebulkbound}. We could also use a different value of $Z$ provided that, in calculating higher point functions, we consistently use the bulk to boundary propagator that comes from taking the limit above.  
 When we take both points to the boundary, we recover the two point function of the boundary operator.
\be
\begin{split}
\langle O(\vect{k}) O(-\vect{k}) \rangle &= Z^2 \lim_{z_2 \rightarrow
  0} z_2^{-\Delta} \lim_{z_1 \rightarrow 0} z_1^{-\Delta} G(\vect{k},
z_1, z_2)  \\ &=  -(2 \Delta - d){\Gamma({d \over 2} + 1 - \Delta) \over \Gamma(\Delta - {d \over 2} + 1)}
   \left({\norm{k_1} \over 2}\right)^{2 \Delta - d} \delta(\vect{k_1}
   + \vect{k_2}) + \ldots.
\end{split}
\ee
Here, once again, as we take $z_2 \rightarrow 0$, we find a divergent term that is analytic in the momenta and so a delta function in position space. This is indicated by the $\ldots$, which are unimportant.  

Note that this prescription is somewhat more straightforward than evaluating the on-shell action, since we don't have to worry about the subleading terms in $\epsilon$ in imposing \eqref{boundcndns} and so we will use it for the stress tensor and conserved currents.

\subsection{Two Point Function of the Stress Tensor and Currents \label{twopointstressgrav}}
To evaluate the two point function of the stress-tensor using AdS/CFT, we simply need to evaluate the two point function of the metric fluctuation in AdS. We will consider the Hilbert-Einstein action:
\be
\label{heaction}
S_{\text{grav}} = {-1 \over 16 \pi G_N} \int \sqrt{-g} \left(R - 2 \Lambda\right),
\ee
where $\Lambda$ is the cosmological constant. We now expand the metric out as:
\be
g_{\mu \nu} = g_{\mu \nu}^{\text{ads}} + h_{\mu \nu}.
\ee

The propagator, in the gauge where we set $h_{z i} = h_{z z} = 0$  is easily evaluated and found to be 
\cite{Christensen:1979iy,Raju:2011mp}:
\begin{equation}
\label{gravitypropagator}
\begin{split}
G^{{\text{grav}}}_{i j, k l}&(\vect{k}, z_1, z_2)  \\ &=
{8 \pi G_N} \int \left[{
z_1^{{d \over 2}-2} J_{{d \over 2}}(p z_1) J_{{d \over 2}} (p z_2) (z_2)^{{d \over 2} - 2} \over  
\left(\vect{k}^2 + p^2 - i \epsilon\right)} \right.   {1 \over 2} \left.\left({\cal T}_{i k} {\cal T}_{j l} + {\cal T}_{i l} {\cal T}_{j k} - 
{2 {\cal T}_{i j} {\cal T}_{k l}\over d-1} \right)\right] {-  d p^2
\over 2 },  
\end{split}
\end{equation}
where ${\cal T}_{i j} = \eta_{i j} + k_{i} k_{j}/p^2$.

First let us take the limit $z_1 \rightarrow 0$, and take $Z$ in \eqref{alternateadscft} to be $Z = -{d \over 8 \pi G_N}$. With this, we see that when we take $z_1 \rightarrow 0$:
\be
\label{gravbulkbound}
Z \lim_{z_1 \rightarrow 0} z_1^{2 - d} G^{{\text{grav}}}_{i j, k l}(\vect{k}, z_1, z_2) =  {1 \over 2} \left({\cal \widetilde{T}}_{i k} {\cal \widetilde{T}}_{j l} + {\cal \widetilde{T}}_{i l} {\cal \widetilde{T}}_{j k} - 
{2 {\cal \widetilde{T}}_{i j} {\cal \widetilde{T}}_{k l}\over d-1} \right)
\left[\frac{2^{\frac{-d}{2}  + 1}}{\Gamma \left(\frac{d}{2}\right)}  \norm{k}^{\frac{d}{2}} z_2^{{d \over 2} - 2} K_{d \over 2} (\norm{k} z_2)  \right],
\ee
where ${\cal \widetilde{T}}_{i j} = \eta_{i j} - k_{i} k_{j}/\norm{k}^2$.

For $d = 3$, which is the case that we are interested in, this takes on a very simple form:
\be
Z \lim_{z_1 \rightarrow 0} z_1^{2 - d} G^{{\text{grav}}}_{i j, k l}(\vect{k}, z_1, z_2)
= {1 \over 2 z_2^2} \left({\cal \widetilde{T}}_{i k} {\cal \widetilde{T}}_{j l} + {\cal \widetilde{T}}_{i l} {\cal \widetilde{T}}_{j k} - 
{{\cal \widetilde{T}}_{i j} {\cal \widetilde{T}}_{k l}} \right) e^{-\norm{k} z_2} (1 + \norm{k} z_2).
\ee
This is the bulk to boundary propagator that we will use below. 

Taking the limit as $z_2 \rightarrow 0$, we now find that:
\be
\langle T_{i j}(\vect{k}) T_{k l}(-\vect{k}) \rangle =   -{1 \over 8 \pi G_N} \norm{k}^d {\Gamma(1 -  {d \over 2} ) \over \Gamma({d \over 2} + 1)} {d \over 2} \left({\cal \widetilde{T}}_{i k} {\cal \widetilde{T}}_{j l} + {\cal \widetilde{T}}_{i l} {\cal \widetilde{T}}_{j k} - 
{2 {\cal \widetilde{T}}_{i j} {\cal \widetilde{T}}_{k l}\over d-1} \right).
\ee
Let us now specialize to the case where $d = 3$. We now have:
\be
\langle T_{i j}(\vect{k}) T_{k l}(-\vect{k}) \rangle =  {4 \over 8 \pi G_N} \norm{k}^3 \left({\cal \widetilde{T}}_{i k} {\cal \widetilde{T}}_{j l} + {\cal \widetilde{T}}_{i l} {\cal \widetilde{T}}_{j k} - {{\cal \widetilde{T}}_{i j} {\cal \widetilde{T}}_{k l}} \right), \quad \text{for~}d=3.
\ee
This matches with the answer obtained from the CFT in (\ref{et}).

Similarly, we can obtain the two point function of currents in the Maxwellian theory in the bulk. (For this, we set $\gamma = 0$, for the moment.) We start with the Maxwell action:
\be
S_{\text{gauge}} = {-1 \over 4 g_4^2} \int \sqrt{-g} F_{\mu \nu} F^{\mu \nu}.
\ee
The bulk to bulk propagator of currents in ``axial gauge'' (where we set the $z$ component of the gauge field to $0$) is given by:
\begin{equation}
\label{axialpropagator}
\begin{split}
&G^{\text{axial},{\rm a  b}}_{i j}(\vect{k}, z_1, z_2) = g_4^2 \int {- d p^2 \over 2 (2 \pi)^d }  \Bigl[{(z_1 z_2)^{\nu_1} J_{\nu_1}(p z_1) J_{\nu_1} (p z_2) {\cal T}_{i j} \delta^{a b}\over 
\left(\vect{k}^2 + p^2 - i \epsilon \right)}\Bigr],
\end{split}
\end{equation}
with $\nu_1 = {d \over 2}  - 1 $.
Repeating the process above and now taking $Z = {2-d \over g_4^2}$, we find that when $z_1 \rightarrow 0$, we get:
\be
Z \lim_{z_1 \rightarrow 0} z_1^{1 - d} G^{{\text{axial}}}_{i j}(\vect{k}, z_1, z_2) = \frac{2^{\frac{1}{2} (2-d ) + 1}}{\Gamma
   \left(\frac{d}{2} - 1\right)}  \norm{k}^{\frac{d}{2} - 1} z_2^{{d \over 2} - 1} K_{{d \over 2} - 1} (\norm{k} z_2) {\cal \widetilde{T}}_{i j}.
\ee
For $d = 3$, we simply have
\be
Z \lim_{z_1 \rightarrow 0} z_1^{1 - d} G^{{\text{axial}}}_{i j}(\vect{k}, z_1, z_2) =  {\cal \widetilde{T}}_{i j} e^{-\norm{k} z_2}, \quad \text{for}~d=3.
\ee

The two point function of currents is given by:
\be
\langle j_i(\vect{k}) j_j(\vect{-k})  \rangle = {1 \over g_4^2} (2 - d){\Gamma(2 -{d \over 2}) \over \Gamma({d \over 2})}
   \left({\norm{k_1} \over 2}\right)^{d - 2} {\cal \widetilde{T}}_{i j}.
\ee
For $d = 3$, we have the remarkably simple expression
\be
\langle j_i(\vect{k}) j_j(\vect{-k})  \rangle = 
   -{1 \over g_4^2} \norm{k_1}  {\cal \widetilde{T}}_{i j},
\ee
which agrees with (\ref{ej}), and fixes $C_J = 1/g_4^2$.

\section{Spinor Helicity Formalism \label{secspinoreview}}

In this appendix, we review the spinor helicity formalism for
correlation functions in 3 dimensional conformal field theories that
was described briefly in section \ref{sec:setting}.  The spinor
helicity formalism adapted to 3-dimensional Lorentzian CFTs is also
described in section 2 of Ref.~\onlinecite{Raju:2012zs}.

In our conventions, the boundary metric is Lorentzian and mostly positive. This means that for two boundary vectors:
\begin{equation}
\vect{k} \cdot \vect{k} = (k_1)^2 + (k_2)^2 - (k_0)^2.
\end{equation}

We use bold-face for vectors but not their components. We use  $i,j$ etc. for boundary spacetime indices and $\mu, \nu$ etc. for bulk spacetime indices. We use $m, n$ etc. to index particle-number. Finally, the components of a momentum vector come with a naturally lowered index. 

Our $\sigma$ matrix conventions are the following
\begin{equation}
\begin{split}
\sigma^0_{\alpha \dot{\alpha}} = \begin{pmatrix}1&0\\0&1\end{pmatrix}, \quad \sigma^1_{\alpha \dot{\alpha}} = \begin{pmatrix}0&1\\1&0\end{pmatrix}, \\
\sigma^2_{\alpha \dot{\alpha}} = \begin{pmatrix}0&-i\\i&0\end{pmatrix}, \quad \sigma^3_{\alpha \dot{\alpha}} = \begin{pmatrix}1&0\\0&-1\end{pmatrix}.
\end{split}
\end{equation}

Given a three momentum $\vect{k} = (k_0, k_1, k_2)$, as we described
in section \ref{sec:setting},  we convert it into 
spinors using
\begin{equation}
k_{\alpha \dot{\alpha}} = k_0 \sigma^0_{\alpha \dot{\alpha}} +  k_1
\sigma^1_{\alpha \dot{\alpha}} +  k_2 \sigma^2_{\alpha \dot{\alpha}} +
i |\vect{k}| \sigma^3_{\alpha \dot{\alpha}} = \la_{\alpha} \lb_{\dot{\alpha}},
\end{equation}
where
\begin{equation}
|\vect{k}| \equiv \sqrt{\vect{k} \cdot \vect{k}} = \sqrt{k_1^2 + k_2^2 - k_0^2}.
\end{equation}
If $\vect{k}$ is spacelike to start with, then the $\sigma^3$ component will be imaginary. 

In components, we have the following expressions for the spinors
\begin{equation}
\begin{split}
\la &= \bigl(\sqrt{k_0 + i |\vect{k}|}, {k_1 + i k_2 \over \sqrt{k_0 + i |\vect{k}|}} \bigr), \\
\lb &= \bigl(\sqrt{k_0 + i |\vect{k}|}, {k_1 - i k_2 \over \sqrt{k_0 + i |\vect{k}|}} \bigr).
\end{split}
\end{equation}

We have the freedom to rescale the spinors by any complex number: $\la \rightarrow \alpha \la,~\lb \rightarrow {1 \over \alpha} \lb$ without changing the momentum.  If we do this for spinors corresponding to an external particle, then this 
rescales the polarization vectors and amplitudes pick up a simple
phase. 

We can raise and lower spinor indices using the $\epsilon$ tensor. We choose the $\epsilon$
tensor to be $i \sigma_2$ for both the dotted and the undotted indices. This means that 
\begin{equation}
\epsilon^{0 1} = 1 = -\epsilon^{1 0},
\end{equation}
and spinor dot products are defined via 
\be
\dotl[\la_1, \la_2] = \epsilon^{\alpha \beta} \la_{1 \alpha} \la_{2 \beta} = \la_{1 \alpha} \la_{2}^{\alpha}, \quad \dotl[\lb_1, \lb_2] = \epsilon^{\dot{\alpha} \dot{\beta}} \lb_{1 \dot{\alpha}} \lb_{2 \dot{\beta}} = \lb_{1 \dot{\alpha}} \lb_2^{\dot{\alpha}}. 
\ee

In the case of four-dimensional flat-space scattering amplitudes, all
expressions can be written in terms of the two kinds of dot products
above. However, in our case, we should expect our expressions for CFT$_3$ correlators to only have a manifest $SO(2,1)$ invariance. This means that we might have mixed products between dotted and undotted indices. Such a mixed product extracts the $z$-component
of vector and is performed by
contracting with $\sigma^3$
\begin{equation}
\label{mixedproduct}
2 i |\vect{k}| = (\sigma^3)^{\alpha \dot{\alpha}} k_{\alpha \dot{\alpha}} \equiv  \dotlm[\la, \lb].
\end{equation}
The reader should note that we use square brackets only for this
mixed product; products of both left and right handed spinors are denoted by angular brackets. Second, we note that this mixed dot product is symmetric:
\begin{equation}
\dotlm[\la, \lb] = \dotlm[\lb, \la].
\end{equation}

When we take the dot products of two 3-momenta, we have
\begin{equation}
\begin{split}
&\vect{k} \cdot \vect{q} \equiv \bigl(k_1 q_1 + k_2 q_2 - k_0 q_0 \bigr) \\ &=  -{1 \over 2} \Big(\dotl[\la_k, \la_q] \dotlb[\lb_k, \lb_q] + {1\over 2} \dotlm[\la_k,\lb_k] \dotlm[\la_q,\lb_q] \Big).
\end{split}
\end{equation}

Another fact to keep in mind is that
\begin{equation}
\begin{split}
&\vect{k_1} + \vect{k_2} = \vect{k_3} \\ &\Rightarrow \la_1 \lb_1 + \la_2 \lb_2 = \la_3 \lb_3 + {1 \over 2} \bigl(\dotlm[\la_1,\lb_1] +  \dotlm[\la_2,\lb_2] - \dotlm[\la_3,\lb_3] \bigr) \sigma^3.
\end{split}
\end{equation}

We also need a way to convert dotted to undotted indices.  We write
\begin{equation}
\lad_{\dot{\alpha}} = \sigma^3_{\alpha \dot{\alpha}} \la^{\alpha}, \quad \lbd_{\alpha} = \sigma^3_{\alpha \dot{\alpha}} \lb^{\dot{\alpha}}.
\end{equation}
This has the property that
\begin{equation}
\dotlb[\mb, \lad] = \dotlm[\mb,\la],
\end{equation}
where the quantity on the right hand side is defined in \eqref{mixedproduct}.

With all this, we can write down polarization vectors for conserved currents. The polarization vectors for a momentum vector $\vect{k}$ associated with
spinors $\la, \lb$ are given by
\begin{equation}
\label{polarizationvects2}
\begin{split}
&\epsilon^+_{\alpha \dot{\alpha}} = 2 {\lbd_{\alpha} \lb_{\dot{\alpha}} \over \dotlm[ \la, \lb]} =  {\lbd_{\alpha} \lb_{\dot{\alpha}} \over i \norm{k}}, \\ 
&\epsilon^-_{\alpha \dot{\alpha}} =  2 {\la_{\alpha} \lad_{\dot{\alpha}} \over \dotlm[\la, \lb]} = {\la_{\alpha} \lad_{\dot{\alpha}} \over i \norm{k}}.
\end{split}
\end{equation}
These vectors are normalized so that 
\begin{equation}
\label{normalizationpol}
\vect{\epsilon^+} \cdot \vect{\epsilon^{+}} = \vect{\epsilon^-} \cdot \vect{\epsilon^{-}} = 0, \quad \vect{\epsilon^+} \cdot \vect{\epsilon^{-}} = 2.
\end{equation}
Polarization tensors for the stress tensor are just outer-products of these vectors with themselves:
\be
e^{\pm}_{i j} = \ep^{\pm}_i \ep^{\pm}_j.
\ee

\bibliographystyle{JHEPmod}
\bibliography{references}
\end{document}